\documentclass[a4paper,11pt]{article}
\pdfoutput=1
\usepackage{jheppub}

\usepackage[utf8]{inputenc}

\usepackage{amssymb}
\usepackage{amsmath}
\usepackage{braket}
\usepackage{xcolor}
\usepackage{graphicx}
\usepackage{slashed}

\DeclareRobustCommand{\Sec}[1]{Sec.~\ref{#1}}

\DeclareRobustCommand{\App}[1]{App.~\ref{#1}}
\DeclareRobustCommand{\Tab}[1]{Table~\ref{#1}}

\DeclareRobustCommand{\Fig}[1]{Fig.~\ref{#1}}

\DeclareRobustCommand{\Eq}[1]{Eq.~(\ref{#1})}

\usepackage{array}
\usepackage{floatrow}
\DeclareFloatFont{tiny}{\tiny}
\newcolumntype{M}{>{$\displaystyle} c <{$}}

\allowdisplaybreaks

\newcommand{\bea}{\begin{eqnarray}}
\newcommand{\eea}{\end{eqnarray}}

\def\Ag4{{A_{g,4}}}
\def\Dg4{{D_{g,4}}}

\newcommand{\e}{\epsilon}

\newcommand{\als}{\alpha_s}

\newcommand{\EEC}{{\text{EEC}}}
\newcommand{\gcusp}{\Gamma^{\mathrm{cusp}}}

\newcommand{\Ord}{\mathcal{O}}

\newcommand{\nn}{\nonumber}

\newcommand{\Gcusp}{\Gamma^{\text{cusp}}}

\definecolor{darkblue}{rgb}{0.0,0.0,0.7}
\newcommand{\CS}{\textcolor{darkblue}}

\def\cF{\mathcal{F}}

\def\cN{\mathcal{N}}
\def\cO{\mathcal{O}}

\def\cP{\mathcal{P}}

\def\cW{\mathcal{W}}

\def\G{\Gamma}

\def\qp{\vec{q}_\perp}

\def\bp{\vec{b}_\perp}
\def\bpsq{\vec{b}_\perp^{\,\,2}}

\newcommand{\abs}[1]{\lvert#1\rvert}

\preprint{}

\title{The Four Loop QCD Rapidity Anomalous Dimension}

\author[a]{Ian Moult,}
\author[b]{Hua Xing Zhu,}
\author[c,1]{and Yu Jiao Zhu}
\note{Corresponding author}

\affiliation[a]{Department of Physics, Yale University, New Haven, CT 06511, USA\vspace{0.5ex}}
\affiliation[b]{Zhejiang Institute of Modern Physics, Department of Physics, Zhejiang University, Hangzhou, 310027, China}
\affiliation[c]{Bethe Center for Theoretical Physics, Universitat Bonn, D-53115, Germany\vspace{0.5ex}}

\emailAdd{ian.moult@yale.edu}
\emailAdd{zhuhx@zju.edu.cn}
\emailAdd{yzhu@uni-bonn.de}

\abstract{The rapidity anomalous dimension controls the scaling of transverse momentum dependent observables in the Sudakov region. In a conformal theory it is equivalent to the soft anomalous dimension, but in QCD this relation is broken by anomalous terms proportional to the $\beta$-function. In this paper we first give a simple proof of this relation using two different representations of the energy-energy correlator observable. We then calculate the anomalous terms to three loops by computing the three-loop fully differential soft function to $\mathcal{O}(\epsilon)$. Combined with recent perturbative data from the study of on-shell form factors and splitting functions, this allows us to derive the four loop rapidity anomalous dimension in QCD.  
}

\begin{document}

\maketitle


\section{Introduction}
\label{sec:introduction}

Precision calculations in perturbative QCD play a crucial role in interpreting data in collider experiments. This is particularly true for key observables, such as QCD event shapes, which can be used for precision measurements of the strong coupling \cite{Becher:2008cf,Abbate:2010xh,Hoang:2015hka,Kardos:2018kqj}, or the Sudakov region of the Higgs transverse moment spectrum, where precision enables study of the coupling of the Higgs to light quarks \cite{Bishara:2016jga,Soreq:2016rae}. Due to progress in understanding the structure and universality of infrared evolution equations, as well as in fixed order calculations, these observables can be computed to next-to-next-to-next-to-leading logarithm (N$^3$LL) matched to next-to-next-to-leading order (NNLO) (see e.g. \cite{Abbate:2010xh,Hoang:2015hka,Bizon:2017rah,Chen:2018pzu,Bizon:2018foh,Cieri:2018oms,Bizon:2019zgf,Bacchetta:2019sam,Kardos:2020gty,Ebert:2020dfc,Becher:2020ugp,Billis:2021ecs,Camarda:2021ict,Re:2021con,Chen:2021vtu,Chen:2022cgv,Ju:2021lah}). 

In the last several years, there has been remarkable progress in extending calculations of a number of key gauge theory quantities to the four loop level. These include the $\beta$-function \cite{Baikov:2016tgj,Herzog:2017ohr,Luthe:2017ttg,Chetyrkin:2017bjc}, the anomalous dimensions of twist-2 operators  \cite{Ruijl:2016pkm,Moch:2017uml,Herzog:2018kwj,Moch:2018wjh}, the cusp anomalous dimension \cite{Grozin:2015kna,Henn:2016men,Ruijl:2016pkm,vonManteuffel:2016xki,Lee:2016ixa,Lee:2017mip,Moch:2018wjh,Henn:2019rmi,Lee:2019zop,vonManteuffel:2019wbj,Bruser:2019auj,Henn:2019swt,vonManteuffel:2020vjv}, the quark and gluon collinear anomalous dimensions \cite{Lee:2016ixa,Henn:2016men,vonManteuffel:2016xki,Lee:2017mip,vonManteuffel:2019wbj,Lee:2019zop,vonManteuffel:2019wbj,Das:2019btv,vonManteuffel:2020vjv,Agarwal:2021zft}, and most recently the full form factors \cite{Lee:2021uqq,Lee:2022nhh,Chakraborty:2022yan}. These anomalous dimensions govern the behavior of scattering amplitudes and cross sections in infrared limits, allowing the resummation of logarithms that appear in these limits, and which often invalidate a fixed order perturbative expansion, to N$^4$LL. They have recently seen their first phenomenological applications with the calculation of the four loop threshold corrections to Higgs production \cite{Das:2020adl}.

However, for a number of the most important phenomenological observables an additional anomalous dimension, the rapidity anomalous dimension, is required to describe the infrared dynamics. This is the case for the energy-energy-correlator (EEC) \cite{Basham:1978bw} in the Sudakov observable, a key observable for precision studies of QCD event shapes, as well as for the transverse momentum, $p_T$, spectrum of the Higgs/electroweak bosons at small $p_T$. More generally, the rapidity anomalous dimension controls the scaling evolution of transverse momentum dependent observables, as formulated in the Collins-Soper-Sterman approach \cite{Collins:1981uk,Collins:1981va,Collins:1984kg} or in the rapidity renormalization group \cite{Chiu:2011qc,Chiu:2012ir} approach in SCET \cite{Bauer:2000ew, Bauer:2000yr, Bauer:2001ct, Bauer:2001yt,Bauer:2002aj}, and is therefore a universal anomalous dimension describing the infrared dynamics of QCD. The goal of this paper is to fill this gap, and compute the four loop rapidity anomalous dimension in QCD. For the EEC, this allows resummation at N$^4$LL in the back-to-back region.\footnote{The five loop cusp anomalous dimension is also required to perform resummation at N$^4$LL, however, its effect should be minuscule. A rough estimate of the value of the five loop cusp can be obtained from the calculation of the five loop non-singlet anomalous dimensions \cite{Herzog:2018kwj}. Only the four loop anomalous dimensions in this paper are required to predict the singular structure at N$^4$LO.} For the case of the $p_T$ spectra of the Higgs or electroweak bosons, the complete four loop DGLAP evolution is also required. This is currently known for the non-singlet splitting functions \cite{Moch:2017uml}, and will hopefully be known for the singlet combinations in the near future.

In this paper we compute the four loop rapidity anomalous dimension in QCD by exploiting its relation in a conformal field theory (CFT) to the soft anomalous dimension, $\gamma_S$. This relation was discovered through perturbative calculations to three loops in \cite{Li:2016ctv}, and subsequently proven to all orders in \cite{Vladimirov:2016dll,Vladimirov:2017ksc}. In a non-conformal theory this relation is broken by terms proportional to the $\beta$-function. The key advantage of exploiting this relation, is that since these corrections multiply $\beta$-functions, in perturbation theory they only need to be computed to one lower order. Therefore, we are able to compute the four loop anomalous dimension using known results for the soft function combined with a three loop calculation.  Technically, we will obtain the anomalous terms by calculating the fully differential soft function \cite{Li:2011zp} to $\cO(\epsilon)$ at three loops in dimensional regularization.\footnote{This approach, as well as the result for the four loop anomalous dimensions were presented by the authors at World SCET 2020~\cite{moult_talk} and at an online QCD seminar hosted by Xiangdong Ji~\cite{zhu_talk}. }

In addition to presenting the result for the four loop rapidity anomalous dimension in QCD, we also give an additional proof of the correspondence between the rapidity anomalous dimension and the soft anomalous dimension in a conformal theory using a recent derivation by Korchemsky of the back-to-back asymptotics of the EEC in a conformal field theory from a representation in  terms of a Euclidean four point function \cite{Korchemsky:2019nzm}. The equivalence of this result with the timelike factorization derived in \cite{Moult:2018jzp} provides a simple proof of this equivalence. We hope that the results of this paper clarify the relations between different anomalous dimensions used in the literature, and emphasizes the importance of understanding relations between them.

An outline of this paper is as follows. In \Sec{sec:summary} we summarize recent progress in the calculation of four loop anomalous dimensions which we will use in our calculation of the four loop anomalous dimension. In \Sec{sec:rap_sum}, we review the definition of the rapidity anomalous dimension as the anomalous dimension of a particular configuration of Wilson lines. In \Sec{sec:proof1} we present a simple proof of the correspondence between soft and rapidity anomalous dimensions in a conformal field theory using two different representations of the Sudakov asymptotics of the energy-energy correlators.  In \Sec{sec:Fully} we describe our calculation of the N$^3$LO fully differential soft function to $\cO(\epsilon)$. In \Sec{sec:SR} we compute the difference between the soft anomalous dimension $\gamma_S$ and the rapidity anomalous dimension $\gamma_R$ at four loops, and hence derive the four loop QCD rapidity anomalous dimension.  We conclude in \Sec{sec:conclusion}. Additional perturbative data is collected in the Appendix, as well as the attached ancillary files.

\section{Summary of Perturbative Data for Anomalous Dimensions}
\label{sec:summary}

In this section we summarize known results for QCD anomalous dimensions that we will use as perturbative data, and discuss some relations between these anomalous dimensions. This will also allow us to state our conventions and notation, since a number of anomalous dimensions are given different names in different contexts.

Apart from the $\beta$-function, the simplest and most well studied anomalous dimensions in gauge theories are the anomalous dimensions of twist-2 spin-$J$ operators, or equivalently, the moments of the spacelike splitting functions.  In the limit $S\to \infty$, the twist-2 anomalous dimensions behave as (see e.g. \cite{Korchemsky:1988si,Alday:2007mf})
\begin{align}\label{eq:high_spin_op}
\gamma(J, \alpha_s)=\gcusp(\alpha_s)( \log J +\gamma_E) - B_\delta (\alpha_s)+\cO(1/J)\,.
\end{align}
This is equivalent to the limit $z\to 1$ in the splitting functions, where they behave as \cite{Korchemsky:1988si,Korchemsky:1992xv,Belitsky:2008mg}
\begin{align}
P(z, \alpha_s)=\frac{\gcusp (\alpha_s)}{(1-z)_+}+ B_\delta (\alpha_s) \delta(1-z) + \cdots \,.
\end{align}
These limits define two anomalous dimensions, the cusp anomalous dimension \cite{Korchemsky:1987wg}, $\gcusp$, which describes the leading divergences for a cusped Wilson line, and the virtual anomalous dimension, $B_\delta$. 

Significant perturbative data is available for both of these anomalous dimensions. In planar $\cN=4$ super Yang-Mills, both $\gcusp$ \cite{Beisert:2006ez,Eden:2006rx} and $B_\delta$ \cite{Freyhult:2007pz,Freyhult:2009my,Fioravanti:2009xt} are known to all loops from integrability. In QCD (and in non-planar $\cN=4$ SYM), the cusp anomalous dimension was recently obtained analytically to four loops \cite{Huber:2019fxe,Henn:2019swt,vonManteuffel:2020vjv} following much earlier work (see \cite{Grozin:2015kna,Henn:2016men,Ruijl:2016pkm,vonManteuffel:2016xki,Lee:2016ixa,Lee:2017mip,Moch:2018wjh,Henn:2019rmi,Lee:2019zop,vonManteuffel:2019wbj,Bruser:2019auj} for the analytic calculation of the planar and matter dependent terms, and \cite{Moch:2017uml,Boels:2017skl,Boels:2017ftb,Boels:2017fng,Moch:2018wjh} for numerical extractions). Due to an extensive program calculating splitting functions in QCD (see e.g. \cite{Ruijl:2016pkm,Moch:2017uml,Herzog:2018kwj,Moch:2018wjh} for recent results), $B_\delta$ is fully known analytically at three loops \cite{Vogt:2004mw,Moch:2004pa} for both quarks and gluons. All color structures in  $B_\delta$ are also known at four loops for both quarks and gluons, either numerically, or analytically. In particular, the planar result is known analytically \cite{Moch:2017uml}, as are results for the $n_f^2$ and $n_f^3$ terms \cite{Gracey:1994nn,Davies:2016jie}. Updated numerical results for all additional color structures were given in \cite{Das:2019btv} for the quark case, and in \cite{Das:2020adl} for the gluon case. For convenience, these anomalous dimensions are summarized in the Appendix.

Our second source of perturbative data will come from the study of on-shell form factors. The logarithms in the on-shell form factors are well studied (see e.g. \cite{Sudakov:1954sw,Collins:1980ih,Sen:1981sd,Korchemsky:1988hd,Collins:1989bt,Magnea:1990zb,Magnea:2000ss,Dixon:2008gr}). Here we follow the notation of \cite{Falcioni:2019nxk}. The all orders formula for a form factor with momentum transfer $Q^2$, evaluated at the scale $\mu^2=Q^2$ can be written 
\begin{align}
F(1,\alpha_s(Q), \epsilon)=\exp \left[  \frac{1}{2} \int_0^{Q^2}  \frac{d\lambda^2}{\lambda^2} \left(  G(1,\alpha_s(\lambda,\epsilon),\epsilon) -\gcusp(\alpha_s(\lambda,\epsilon)) \log\frac{Q^2}{\lambda^2} \right) \right]\,,
\end{align}
where $\epsilon = 2 - d/2$ is the dimensional regulator, and $\alpha_s(\lambda, \epsilon)$ is the $d$-dimensional running coupling at scale $\lambda$.
As with the twist-2 operators, the leading logarithms are governed by the cusp anomalous dimension. The subleading logarithms are governed by the so called ``collinear anomalous dimension", $\gamma_G$, which is defined by
\begin{align}
\int_0^{\mu^2} \frac{d\lambda^2}{\lambda^2}G(1,\alpha_s(\lambda,\epsilon),\epsilon)=\int_0^{\mu^2} \frac{d\lambda^2}{\lambda^2} \gamma_G\big(\alpha_s(\lambda,\epsilon)\big) +\cO(\epsilon^0)\,,
\end{align}
where it is understood that all the poles are absorbed by the integration with $\gamma_G$.
The collinear anomalous dimension is known analytically to four loops in planar $\cN=4$ SYM \cite{Dixon:2017nat} (for an earlier numerical calculation, see \cite{Cachazo:2007ad}), and has been computed numerically in non-planar $\cN=4$ SYM \cite{Boels:2017ftb,Boels:2017fng}. In QCD there has been an extensive program computing the four loop form factors. In particular, it was computed in the planar limit in \cite{Lee:2016ixa,Henn:2016men}, the $n_f^3$ terms were computed in \cite{vonManteuffel:2016xki}, the $n_f^2$ terms were computed in \cite{Lee:2017mip,vonManteuffel:2019wbj} and quartic color factor contributions were computed in \cite{Lee:2019zop,vonManteuffel:2019wbj,Das:2019btv}. Very recently all matter dependent terms were obtained in \cite{vonManteuffel:2020vjv} (using previous results e.g. \cite{vonManteuffel:2015gxa,vonManteuffel:2019gpr}), and finally the full four loop form factors were computed \cite{Lee:2021uqq,Lee:2022nhh,Chakraborty:2022yan}.

The anomalous dimension that we are ultimately interested in, and which is closely related to the rapidity anomalous dimension is the ``soft anomalous dimension", $\gamma_S$, which can be expressed in terms of the anomalous dimensions described above as \cite{Dixon:2008gr,Ravindran:2004mb,Falcioni:2019nxk}
\begin{align}\label{eq:soft_define}
\gamma_G-2B_\delta=f_{\text{eik}}= -2 \gamma_S\,.
\end{align}
The relation connecting collinear, virtual, and soft anomalous dimension, \eqref{eq:soft_define}, can be easily understood by considering threshold resummation using SCET, where all these three anomalous dimension appear, see e.g.~\cite{Manohar:2003vb,Idilbi:2005ky,Idilbi:2005ni,Becher:2007ty}.
The soft anomalous dimension is sometimes also referred to as $f_{\text{eik}}$, since it can be thought of as an eikonal version of the collinear anomalous dimension. For a recent discussion of this relation and other similar relations, see \cite{Falcioni:2019nxk}. This anomalous dimension controls soft gluon radiation from Wilson lines, as we will describe in more detail in \Sec{sec:rap_sum}, and describes for example, Drell-Yan or gluon fusion at threshold, or dijet event shapes in the back-to-back limit, see e.g. \cite{Moch:2005tm,Becher:2007ty,Li:2014afw}.

Using the known perturbative data for $\gamma_G$, and $B_\delta$, the soft anomalous dimensions were assembled using \Eq{eq:soft_define} in \cite{Das:2019btv,Das:2020adl} (additional data was recently provided in \cite{vonManteuffel:2020vjv} has been incorporated in our results). This provides the following result for the four loop soft anomalous dimension in QCD:
\begin{align}\label{eq:3soft}
     \gamma^S_{3} =& 
 \CS{C_A^3 C_r }
 \bigg(-\frac{b(d^{(4)}_{FA})}{24}
 -\frac{9311591}{13122}+\frac{2189 }{9}\zeta _3^2+\bigg(\frac{1057}{3} \zeta _4+\frac{414602}{243}\bigg) \zeta _3-\frac{53467}{27}\zeta _5
 \nn\\
 -&\frac{115759}{54}\zeta _4 +\frac{10868}{9}\zeta _6 +\frac{11071}{12}\zeta _7+\bigg(-\frac{6284}{9}\zeta _3+\frac{688}{3}\zeta _5+\frac{1164703}{1458}\bigg) \zeta _2\bigg)
 \nn\\
+&\CS{N_f  C_A^2 C_r} \bigg(-\frac{1}{2} b\bigg( N_f C_F C_A C_r\bigg)-\frac{1}{4} b\bigg(N_f C_F^2 C_r \bigg)-\frac{b(d^{(4)}_{FF})}{48}-\frac{3506 \zeta _3^2}{9}-\frac{5615}{18}\zeta _6
\nn\\
-&\frac{74015}{243}\zeta _3-\frac{5276}{27}\zeta _5+\frac{29059}{27}\zeta _4+\bigg(\frac{1208}{9}\zeta _3-\frac{160729}{729}\bigg) \zeta _2+\frac{130625}{3888}\bigg)
\nn\\
+&\CS{N_f  C_A C_F C_r} \bigg(b\bigg(N_f C_F C_A C_r \bigg)+\frac{1700 \zeta _3^2}{3}-\frac{26372}{27}\zeta _4+\frac{11759}{81}\zeta _3+\frac{2236}{3}\zeta _5-\frac{1952}{9}\zeta _3 \zeta _2
\nn\\
-&\frac{673}{54} \zeta _2 -359 \zeta _6+\frac{1092511}{1944}\bigg)+ \CS{N_f  C_F^2 C_r} \bigg(b\bigg(N_f C_F^2 C_r \bigg)-184 \zeta _3^2-\frac{1936}{3}\zeta _5+\frac{560}{9}\zeta _3
\nn\\
+&\frac{7334}{9}\zeta _6+\bigg(\frac{256}{3}\zeta _3-161\bigg) \zeta _2+167 \zeta _4-\frac{27949}{216}\bigg)+ \CS{C_A C_r N_f^2} \bigg(-\frac{16076}{243}\zeta _3-\frac{194}{9}\zeta _4
\nn\\
+&\frac{112}{9}\zeta _3 \zeta _2
+\frac{15481}{1458}\zeta _2+56 \zeta _5-\frac{27875}{34992}\bigg)
+\CS{C_F C_r N_f^2} \bigg(-\frac{152}{9}\zeta _5-\frac{32}{9}\zeta _4
+\frac{2284}{81}\zeta _3-\frac{16}{3}\zeta _3 \zeta _2
\nn\\
+&\frac{86}{9} \zeta _2
-\frac{16733}{972}\bigg)  + \CS{C_r N_f^3}\bigg(-\frac{64}{27}\zeta _4+\frac{8}{81}\zeta _2
+\frac{200}{243}\zeta _3+\frac{8080}{6561}\bigg)\,,
\nn\\
+ & 
     \CS{\frac{ d_A^{abcd} d_r^{abcd} }{N_r}}
 \bigg(b(d^{(4)}_{FA})+\frac{1672 \zeta _3^2}{3}+\bigg(184 \zeta _4+\frac{3904}{9}\bigg) \zeta _3-\frac{1738}{9}\zeta _6 -1742 \zeta _7-96
 \nn\\
 -&\frac{56}{3}\zeta _4+\frac{920}{9}\zeta _5+\zeta _2 \bigg(896 \zeta _3-512 \zeta _5+\frac{1088}{3}\bigg)\bigg)
 \nn
 \\
 &
 +\CS{N_f \frac{d_r^{ abcd}d_F^{ abcd} }{N_r}}\bigg(b(d^{(4)}_{FF})-\frac{608 }{3}\zeta _3^2-\frac{592}{9}\zeta _6+\frac{400}{3}\zeta _4+\frac{2656}{9}\zeta _3 +192
\nn\\
+&\frac{10880}{9}\zeta _5+\zeta _2 \bigg(-64 \zeta _3-\frac{2272}{3}\bigg)\bigg) \,,
\end{align}
where $C_r = C_F$ for fundamental Wilson lines and $C_r = C_A$ for adjoint Wilson lines, and $N_r$ is the dimension of representation, $N_r = N_c$ for fundamental and $N_r = N_c^2 - 1$ for adjoint. We have defined here and below a perturbative expansion for a generic anomalous dimension $\gamma(\alpha_s)$ as
\begin{equation}
\gamma(\alpha_s) = \sum_{i=0}^\infty \left(\frac{\alpha_s}{4 \pi} \right)^{i+1} \gamma_i \equiv  \sum_{i=0}^\infty a_s^{i + 1} \gamma_i  \,.
\end{equation}

The $n_f C_F^2 C_r$ and $n_f C_F C_A C_r$ and all other coefficients in the collinear anomalous dimension were computed analytically (or with high numerical precisions ) in \cite{vonManteuffel:2020vjv,Agarwal:2021zft},
and by combing all the knowledges for the anomalous dimensions,
especially from the  study of $B_\delta$ in ~\cite{Das:2019btv,Das:2020adl},
 it turns out that our results only need four of the numeric values collected in \rm{Table 1} of the work~\cite{Das:2019btv}.
The extra numeric values are linearly dependent on those in \Tab{tab:Numeric_soft}, 
and they are numerically consistent  as compared  with the numbers collected in \rm{Table 1} of  ~\cite{Das:2019btv}.

At four loops, strict Casimir scaling is violated for both the cusp anomalous dimension and the soft anomalous dimension \cite{Armoni:2006ux,Alday:2007mf,Boels:2017skl,Boels:2017ftb,Boels:2017fng,Moch:2017uml,Grozin:2017css,Henn:2019rmi,Lee:2019zop,vonManteuffel:2019wbj}. However a generalized Casimir scaling holds for the cusp, and this is inherited by the soft anomalous dimension, since they exhibit the same maximally non-abelian structure \cite{Ravindran:2004mb,Moch:2005tm,Dixon:2008gr}. The contributions that break naive Casimir scaling are due to the presence of the quartic Casimirs \cite{vanRitbergen:1997va}
\begin{align}
d^{(4)}_{FF}&=\frac{(N_c^4-6N_c^2+18)(N_c^2-1)}{96 N_c^2}\,, \nn \\
d^{(4)}_{FA}&=d^{(4)}_{AF}=\frac{N_c(N_c^2+6)(N_c^2-1)}{48}\,, \nn \\
d^{(4)}_{AA}&=\frac{N_c^2(N_c^2+36)(N_c^2-1)}{24}\,.
\end{align}
The coefficients of these terms in the cusp and form factors were computed in \cite{Moch:2018wjh,Lee:2019zop,Henn:2019rmi}. The coefficients of these quartic Casimirs in both the cusp and eikonal anomalous dimensions exhibit a generalized Casimir scaling, namely that the contributions to the eikonal anomalous dimension from quartic Casimirs are determined by only two functions, instead of three \cite{Moch:2018wjh,Becher:2019avh}. We have employed the generalized Casimir scaling to present the soft function anomalous dimension for generic SU($N_c$) representation $r$ in \eqref{eq:3soft}. We will also exploit this relation later in this paper to obtain the rapidity anomalous dimensions for both quarks and gluons.

\begin{table}[t!]
  \vspace*{-1mm}
  \centering
  \renewcommand{\arraystretch}{1.2}
  \begin{tabular}{MMMMM}
    \hline \hline
     b(N_f\, C_F^2 C_r)          &
    b(N_f\, C_F C_A  C_r)       &
     b(d^{(4)}_{FF})        &
     b(d^{(4)}_{FA})
    \\ \hline
    80.780 \pm 0.005 & 
    -455.247 \pm 0.005 & 
       -143.6 \pm 0.2 & 
    -998.0 \pm 0.2   
\\ \hline
  \end{tabular}
  \vspace*{1mm}
  \label{tab:Numeric_soft}
    \caption{\small{Numerical values for the color coefficients appearing in the soft anomalous dimension.  These are extracted from the $B_\delta$, and are taken from \cite{vonManteuffel:2020vjv}.
    Note that these numeric numbers are independent of representation.   }}
    
  \end{table}

\section{The Rapidity Anomalous Dimension}
\label{sec:rap_sum}

\begin{figure}
  \centering
  \includegraphics[width=0.5\textwidth]{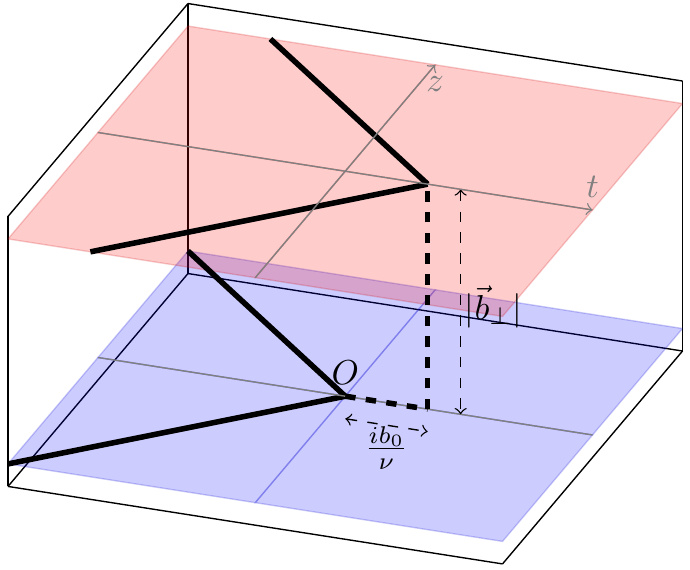}
  \caption{The configuration of Wilson lines used to define the rapidity anomalous dimension. This configuration appears in the description of the EEC in the back-to-back limit, and a related configuration appears in the description of the $p_T$ spectrum.}
  \label{fig:EEC_def}
\end{figure}

The rapidity anomalous dimension governs the rapidity evolution of TMD quantities. Its calculation typically requires a an definition of rapidity regulator. Here we adopt the exponential regulator~\cite{Li:2016axz}, which has been successfully applied at three-loop order~\cite{Li:2016ctv,Luo:2019szz,Ebert:2020yqt,Luo:2020epw,Ebert:2020qef}. Other definitions  of rapidity regulator can be found in \cite{Ji:2004wu,Collins:2011zzd,Becher:2011dz,Chiu:2011qc,Chiu:2012ir,Echevarria:2015byo,Ebert:2018gsn,Vladimirov:2020umg}.  We can define the following matrix element of soft Wilson lines~(in the following we use fundamental Wilson lines as explicit example, while our conclusion apply equally to adjoint Wilson lines.)
\begin{align}\label{eq:soft_eec}
S_\EEC (\vec b_\perp, \mu, \nu) =\lim_{\nu\to +\infty} \frac{1}{N_c} \mathrm{Tr} \langle 0 | T \left[  S^\dagger_{n }(0) S_{\bar n}(0) \right] \bar T \left[ S^\dagger_{\bar n}\left(y_\nu(\vec b_\perp)\right) S_{n}\left(y_\nu(\vec b_\perp)\right)  \right] |0 \rangle\,,
\end{align}
where
\begin{align}
y_\nu (\vec b_\perp) = ( i b_0 / \nu, i b_0/ \nu, \vec b_\perp)\,, \qquad b_0=2e^{-\gamma_E} \,,
\end{align}
and the soft Wilson line is defined as
\begin{equation}
S_n(x) = P \exp \left( i g_s \int_0^{\infty} \! ds \, n \cdot A_s(x + sn) \right) \,.
\end{equation}
 This configuration is shown in \Fig{fig:EEC_def}. This particular matrix element describes soft gluons in transverse momentum resummation, and certain $e^+e^-$ dijet event shapes in the back-to-back limit. Here we have chosen the orientation of the Wilson lines as appropriate for the case of an $e^+e^-$ event shape. For the case of $q_T$, the orientation of the Wilson lines is flipped, but this does not change the associated anomalous dimensions~\cite{Zhu:2020ftr,Catani:2021kcy}, at least through to three loops.

This matrix element exhibits divergences that are regulated by dimensional regularization, leading to an RG equation in $\mu$, and divergences that are regulated by $\nu$, leading to a rapidity renormalization group in $\nu$. The RG equation in $\mu$ is given by
\begin{align}
\mu \frac{dS_\EEC(\vec b_\perp, \mu, \nu)}{d\mu} = \left[  2 \Gamma_{\text{cusp}}(\alpha_s) \ln \frac{\mu^2}{\nu^2} -4 \gamma_S (\alpha_s) \right] S_\EEC (\vec b_\perp, \mu, \nu)\,,
\end{align}
and is governed by the cusp anomalous dimension and the soft anomalous dimension, $\gamma_S$ described above. The evolution in $\nu$ is given by~\cite{Chiu:2011qc,Chiu:2012ir,Li:2016ctv}~\footnote{Note that the convention for $\gamma_R$ is such that it differs from $\gamma_r$ in \cite{Li:2016ctv} by a normalization, $2 \gamma_R =  \gamma_r$. Similarly, the soft anomalous dimension also differed by a factor of $2$ compared with $\gamma_s$ in \cite{Li:2016ctv}, $2 \gamma_S = \gamma_s$. }
\begin{align}\label{eq:nu_RG_S}
\nu \frac{dS_\EEC(\vec b_\perp, \mu, \nu)}{d\nu}= \left[ - 2 \int\limits_{b_0^2/\vec b_\perp^2}^{\mu^2} \frac{d\bar\mu^2}{\bar\mu^2} \Gamma_{\text{cusp}}(\alpha_s(\bar\mu))  + 4 \gamma_R(\alpha_s(b_0/|\vec b_\perp|))  \right] S_\EEC(\vec b_\perp, \mu, \nu)\,,
\end{align}
which is governed by the cusp anomalous dimension, and the rapidity anomalous dimension $\gamma_R$. Identical evolution equations hold for the case of transverse momentum resummation. The goal of this paper will be to compute the rapidity anomalous dimension to four loops in QCD, which completes the scale understanding of this matrix element at four loops.

This soft function was computed to three loops in \cite{Li:2016ctv}, and there it was noticed that in a conformal theory one has the relationship
\begin{equation}
\label{eq:raw_rapid}
\gamma_R = \gamma_S \,.
\end{equation} Note that the rapidity anomalous dimension,  begin a evolution equation in rapidity, contains higher order terms in $\e$, which are neglected in \eqref{eq:raw_rapid}. This relation was subsequently generalized to all orders in \cite{Vladimirov:2016dll,Vladimirov:2017ksc}, where the relation
\begin{align}
\gamma_{R}(\e^*)=\gamma_{S}
\end{align}
was obtained, with explicit dependence of $\e$ replaced by $\e^*$, which is determined by equation $\beta^*(\e^*)\equiv\frac{1}{\als}\beta(\als)-2\e^*=0$ order-by-order in QCD. In other words, it is the critical dimension where QCD is conformal order by order in perturbation theory. This relationship allows us to compute $\gamma_R$ at a given loop order by knowing $\gamma_S$ at that loop order, and $\gamma_R$ to higher orders in $\e$ at lower loop orders. This  approach has been utilized in \cite{Vladimirov:2016dll} to reproduce the three-loop $\gamma_R$ from a two-loop calculation and knowledge of the three-loop $\gamma_S$.  We will adopt this approach in this paper.
For a recent formal study  of the uses of conformal symmetry in $d=4-2\epsilon$, see \cite{Braun:2018mxm}.

\section{Proof of Soft-Rapidity Correspondence Using Energy Correlators}
\label{sec:proof1}

Since we will exploit heavily the correspondence between the rapidity anomalous dimension and the soft anomalous dimension in a CFT, in this section we give an independent proof of this fact using the energy-energy correlator (EEC) observable. This proof exploits the universality of factorization, combined with the fact that the EEC also admits a representation in terms of a Euclidean four point function. By comparing two distinct results for the EEC we are able to prove $\gamma_R=\gamma_S$ in a conformal theory.

The EEC admits a timelike factorization formula \cite{Moult:2018jzp} derived in SCET \cite{Bauer:2000ew, Bauer:2000yr, Bauer:2001ct, Bauer:2001yt,Bauer:2002aj}
\begin{align}
  \label{eq:resformula}
\frac{d\sigma}{dz} = &\; \frac{1}{4} \int\limits_0^\infty db\, b
  J_0(bQ\sqrt{1-z})H(Q,\mu_h) j^q_\EEC(b,b_0/b,Q) j^{\bar
  q}_\EEC(b,b_0/b,Q) S_\EEC( b,\mu_s, \nu_s) 
\nn\\
&\; \cdot
  \left(\frac{Q^2}{\nu_s^2}\right)^{2 \gamma_R(\alpha_s(b_0/b))}
  \exp \left[ \int\limits_{\mu_s^2}^{\mu_h^2}
  \frac{d\bar{\mu}^2}{\bar{\mu}^2}  \Gamma_{\text{cusp}} (\alpha_s(\bar \mu)) \ln
  \frac{b^2\bar{\mu}^2}{b_0^2} \right.
\nn\\
&\;
\left. +
  \int\limits_{\mu_h^2}^{b_0^2/b^2}\frac{d\bar{\mu}^2}{\bar{\mu}^2}
  \left(  \Gamma_{\text{cusp}} (\alpha_s(\bar \mu)) \ln\frac{b^2 Q^2}{b_0^2} +
   \gamma_H (\alpha_s(\bar \mu)) \right) -
  \int\limits_{\mu_s^2}^{b_0^2/b^2}\frac{d\bar{\mu}^2}{\bar{\mu}^2}
   \gamma_S (\alpha_s(\bar \mu))  \right]\,,
\end{align}
where $\gamma_H = - \gamma_G = -2 B_\delta + 2 \gamma_S$ is the collinear anomalous dimension. This factorization formula is identical in structure to that for $p_T$ resummation. The latter has been proven rigorously, including the cancellation of Glauber gluons  \cite{Collins:1988ig}. It has been used to compute the EEC at N$^3$LL' \cite{Ebert:2020sfi} using results for the three-loop transverse momentum dependent fragmentation functions \cite{Ebert:2020qef,Luo:2020epw}. For other studies of the EEC in the back-to-back limit, see  \cite{Gao:2019ojf,Moult:2019vou,Li:2021txc,Li:2020bub}.

This formula applies in both conformal and non-conformal gauge theories, however, to prove the equality of the soft and rapidity anomalous dimensions, we can consider this formula in the conformal limit.  After setting all scales equal to their canonical values such that all large logarithms are absorbed into the Sudakov exponent,
\begin{equation}
\mu_h = Q \,, \qquad \mu_s = \frac{b_0}{b} \,, \qquad \nu_s = \frac{b_0}{b} \,,
\end{equation}
and using the fact that the coupling constant does not run in a conformal theory, we have
\begin{align}
  \label{eq:resformula_N4}
\frac{d\sigma}{dz} = &\; \frac{1}{4} \int\limits_0^\infty db\, b
  J_0(bQ\sqrt{1-z})H(Q,\mu_h) j^q_\EEC(b,b_0/b,Q) j^{\bar
  q}_\EEC(b,b_0/b,Q) S_\EEC( b,\mu_s, \nu_s) 
\nn\\
&\exp \left[-\frac{1}{2}  \Gamma_{\text{cusp}} \log^2\left( \frac{b^2 Q^2}{b_0^2} \right)  +2 B_\delta \log\left( \frac{b^2 Q^2}{b_0^2} \right)  +2 (\gamma_R-\gamma_S)\log\left( \frac{b^2 Q^2}{b_0^2} \right) \right]\,.
\end{align}

The EEC is special in that it also admits a representation as a four point Wightman function \cite{Hofman:2008ar,Belitsky:2013bja,Belitsky:2013xxa}, enabling it to be studied using a completely different set of techniques from conformal field theory. In particular, the EEC can be written as
\begin{align}
\EEC\sim \int d^4x e^{ix\cdot q}  \langle 0 |J_\mu(0) \mathcal{E}(\vec n_1)  \mathcal{E}(\vec n_2) J_\mu(x) | 0 \rangle\,,  
\end{align}
where
\begin{align}\label{eq:ANEC_op}
\mathcal{E}(\vec n) =\int\limits_0^\infty dt \lim_{r\to \infty} r^2 n^i T_{0i}(t,r \vec n)\,,
\end{align}
is an energy flow operator \cite{Sveshnikov:1995vi,Korchemsky:1999kt,Lee:2006nr,Hofman:2008ar,Belitsky:2013xxa,Belitsky:2013bja,Belitsky:2013ofa}.
It is therefore expressible as an integral kernel acting on the four point function of stress tensors. Using an understanding of the behavior of the four point function as operators become lightlike separated \cite{Alday:2010zy,Alday:2013cwa}  and the high spin limit of the twist two operators \Eq{eq:high_spin_op}, Korchemsky proved in \cite{Korchemsky:2019nzm} that the EEC in a conformal theory is governed by the anomalous dimensions $\gcusp$ and $B_\delta$ only, and is described by the formula~\footnote{See e.g. Eq.~(1.5) of \cite{Korchemsky:2019nzm}. We have converted it to the conventions used in this paper.}
\begin{align}
  \label{eq:resformula_N4_korchemsky}
\frac{d\sigma}{dz} = &\; \frac{1}{4} \int\limits_0^\infty db\, b
  J_0(bQ\sqrt{1-z})H(Q,\mu_h) j^q_\EEC(b,b_0/b,Q) j^{\bar
  q}_\EEC(b,b_0/b,Q) S_\EEC( b,\mu_s, \nu_s) 
\nn\\
&\exp \left[-\frac{1}{2}  \Gamma_{\text{cusp}} \log^2\left( \frac{b^2 Q^2}{b_0^2} \right)  + 2 B_\delta \log\left( \frac{b^2 Q^2}{b_0^2} \right) \right]\,.
\end{align}
By comparing this formula derived from the four point function with the timelike factorization formula in \Eq{eq:resformula_N4}, we immediately prove $\gamma_R=\gamma_S$ in a CFT. We therefore arrive at a transparent proof of the relation between soft and rapidity anomalous dimensions.\footnote{It is interesting to note that non-trivial relations between anomalous dimensions (in particular the Basso-Korchemsky reciprocity \cite{Basso:2006nk}) can also be derived by studying equivalent expressions for the EECs in the collinear limit. Ref.~\cite{Chen:2020uvt} provides an explicit example of such relation at three loops in QCD. On the one hand they can be expressed in terms of spacelike anomalous dimensions using the light-ray OPE \cite{Hofman:2008ar,Kologlu:2019mfz,Korchemsky:2019nzm}, and on the other hand they can be expressed in terms of timelike anomalous dimensions using factorization formulas derived in SCET \cite{Dixon:2019uzg,Chen:2020vvp}. These relations suggest much is still to be understood about Lorentzian dynamics.}

Although this relation is sufficient for our purposes, it would be interesting to extend this to study the behavior of rapidity anomalous dimensions for soft functions that have Wilson lines in greater than two directions. Higher point energy correlators exhibit Sudakov behavior in particular kinematic limits, and this has interesting applications at hadron colliders \cite{Gao:2019ojf}. These singular regions can be accessed also on the correlator side. However, unlike the more well studied case \cite{Alday:2010zy}, for three-point correlators and higher the operators become lightlike separated, but not in a sequential limit.

\section{The N$^3$LO Fully Differential Soft Function to $\cO(\epsilon)$}
\label{sec:Fully}

In this section we describe the methods used to compute the N$^3$LO fully differential soft function to $\cO(\epsilon)$ at three loops, from which we will extract the rapidity anomalous dimension, retaining higher-order dependence in $\epsilon$. The fully differential soft function was introduced in~\cite{Li:2011zp} for the study of transverse momentum resummation. It has been used in~\cite{Zhu:2014fma} as a deformation of the threshold soft function, and in \cite{Li:2016axz} as a deformation of TMD soft function. It has also been exploited to calculate Sudakov radiator for jet observable~\cite{Banfi:2018mcq}. It is mostly conveniently defined on the exponent,
\begin{align}
\cF(q)  \equiv \int \prod_i \frac{d^d q_i}{ (2 \pi)^{d-1}} \,  \delta_+(q_i^2) (2 \pi)^d \delta^{(d)}(q-\sum_i q_i) \cW_{n \bar n}(\{q_i\})\,,
\end{align}
where additional symmetry factor for identical final-state particle is implicitly understood. 
The integrand $\cW_{n \bar n}(\{q_i\})$ corresponds to single or multiple soft emissions \emph{web diagrams} from two lightlike Wilson lines $S_n$ and $S_{\bar n}$, such that non-abelian exponentiation theorem is made manifest whenever the observable have that property. Definition of web diagrams can be found, e.g., in \cite{Gatheral:1983cz,Frenkel:1984pz}. The main task of the work here is to calculate the NNNLO fully differential soft function to higher orders in $\e$. The TMD soft function~\cite{Li:2016ctv} and the threshold soft function~\cite{Li:2014afw,Li:2014bfa,Li:2013lsa,Duhr:2013msa} can then all be obtained as limits of the fully differential soft function.

To perform the calculation, Feynman diagrams were generated by \texttt{Qgraph}~\cite{Nogueira:1991ex}~\footnote{We note that several results for the soft web diagrams are available in the literature, namely the two-loop single soft current~\cite{Li:2013lsa,Duhr:2013msa}, one-loop double soft current~\cite{Zhu:2020ftr,Catani:2021kcy}, and tree-level triple soft current~\cite{Catani:2019nqv}. Furthermore, three-loop threshold function has also been obtained in \cite{Anastasiou:2013srw,Anastasiou:2014vaa,Li:2014afw,Li:2014bfa,Anastasiou:2015yha}. We choose to generate the relevant soft currents using an independent code, therefore providing additional checks on the existing results.}, color/Dirac algebra and integrand manipulations were performed in \texttt{form}~\cite{Ruijl:2017dtg}. With the help of reverse unitarity~\cite{Anastasiou:2002yz}, the integrals can  be reduced with Integration-By-Parts identities~\cite{Chetyrkin:1981qh}. We use the publicly available Mathematica package \texttt{LiteRed}~\cite{Lee:2012cn}. We obtained systems of master integrals to formulate differential equations with respect to $n$ and $\bar n$ rescaling invariant $x$, 
\begin{align}
\label{eq:sca_invariant}
x\equiv\frac{q^2 \, (n\cdot\bar n) }{2 (q\cdot n) \, (q\cdot \bar n)} = \frac{q^2}{q^+ q^-} \,,
\end{align}
which can be solved using the method of differential equation~\cite{Kotikov:1990kg}. Note that the final-state total momentum satisfy $q^2 \geq 0$, $x \in [0, 1]$.

The differential equations allow an $\e$-form~\cite{Henn:2013pwa} with the help of \text{CANONICA}~\cite{Meyer:2016slj}, after which we can derive the boundary constants by considering the limit $x\to 1$ and $x\to 0$. 

To all orders in perturbation, the unnormalized results can be written as
\begin{equation}
\cF^{\rm b}(q, \e) =  \sum_{j=0}^{\infty} (a_s^{\rm b})^{j+1} \frac{ F^{T, \rm b}_j (x, \e)}{ (\vec q_\perp\,^2)^{2 + j \e}} \,,
\end{equation}
where the variable $x$ is defined in \eqref{eq:sca_invariant} and $a_s^{\rm b} = \alpha_s^{\rm b}/(4 \pi)$ is the bare coupling constant. We have retain the higher order in $\e$ dependence in the definition of full differential soft function, in anticipating that further integration of $q$ can leads to poles in $\e$. The bare perturbation results for $F^{T, \rm b}(x, \e)$, as well as the renormalized ones,  are given in the ancillary file \texttt{fully.m} through to three loops.

\section{Soft and Rapidity Anomalous Dimensions at N$^4$LO}
\label{sec:SR}

In this section we present the main result of this paper, namely the four loop relation between the soft and rapidity anomalous dimensions in QCD obtained by computing the fully differential soft function at three loops to higher orders in $\epsilon$. We then combine this with the perturbative data summarized earlier to give the four loop rapidity anomalous dimension in QCD.

The bare TMD soft function with renormalized coupling is obtained by perform Fourier transformation and Laplace transformation to the fully differential soft function, and taking the rapidity regulator $\tau = 1/\nu \to 0$,
\begin{align}
\label{eq:baresoft}
s^{\text{b}}(\bp, \nu=1/\tau,\mu)\equiv& \int d^d q \exp\left(-(q\cdot n + q\cdot \bar n) \tau e^{-\gamma_E}-i \bp\cdot\qp \right) \cF^{\rm b}(q, \e)
\nn\\
=&\sum_{j=0} \left(\frac{{\als(\mu)}}{4\pi}\right)^{j+1}\int_0^{1} dx \,\cP_j\left(x,\abs{\bp}\mu/b_0,\mu/\nu\right) \, F^{T, \rm b}_j(x,\e)(1-x)^{-2}\,,
\end{align}
where 
\begin {align}
\cP_j\left(x,\abs{\bp}\mu/b_0,\mu/\nu\right)\equiv& \left(\frac{\tau^2 \mu^2}{b_0^2}\right)^{2+(j+1)\e} \frac{\G(-(j+1)\e )^2}{\G(1-\e)} (1-x)^{2} 
\nn\\
\times&{_2}F_1\left(-(j+1)\e,-(j+1)\e;1-\e;-\frac{\bp^2(1-x)}{\tau^2}\right)\,
\nn\\
\overset{\tau{ \rightarrow 0}}{=}&\left(\frac{\bp\,^2 \mu^2}{b_0^2}\right)^{(j+1)\e}
\frac{{\G(-(j+1)\e)}}{\G(1+j\e)}\bigg[
L_r+\ln(1-x)
\nn\\
 -&{H(-1-(j+1)\e)-H(j\e) }\bigg]\,,
 \label{eq:P}
\end{align}
and $L_r = \ln \left(\nu^2 {\bpsq}/b^2_0\right)$ is the rapidity logarithm, $H$ are the harmonic numbers,
and $F^{T, \rm b}_j(x,\e)$ are the expansion coefficients of the unrenormalized fully differential soft function. The gamma function $\Gamma( -(j+1) \e)$ in \eqref{eq:P} leads to at most double poles in $\e$ for $s^{\rm b}$, which should be renormalized in $\overline{\text{MS}}$ scheme. After operator and coupling constant renormalization, the soft function still contains higher order in $\e$ dependence. Note that the soft function calculated above is really the logarithm of TMD soft function. The usual TMD soft function can be written as
\begin{equation}
S(\vec{b}, \nu, \mu, \e) = \exp( s(\vec{b}, \nu, \mu, \e) ) \,.
\end{equation}

With the renormalized, finite TMD soft function $s(\vec{b}_\perp, \nu, \mu, \e)$, we obtain the rapidity anomalous from 
\begin{align}
\gamma^{R}(\e)\equiv& \left. \frac{1}{2} \frac{\partial s\left(\vec{b}_\perp, \nu, \mu, \e\right)}{\partial L_r} \right|_{L_b\to0}\,.
\end{align}
The explicit expression for $\gamma^{R}(\e)$ can be read off of the results given in \App{sec:TMDsoft}
\begin{align}
\gamma^{R}(\e)=&\frac{\als}{4 \pi} \bigg\{\CS{C_r} \bigg(-\frac{9}{8} \zeta _4 \epsilon ^3-\frac{2}{3} \zeta _3 \epsilon ^2-\zeta _2
   \epsilon \bigg)\bigg\}
   \nn\\
   +&\left(\frac{\als}{4 \pi}\right)^2\bigg\{\CS{C_A C_r} \bigg(\bigg(\frac{46}{3} \zeta _3 \zeta _2-\frac{404 }{27}\zeta _2-\frac{134
   }{27}\zeta _3-\frac{55 }{24}\zeta _4+62 \zeta _5-\frac{14576}{243}\bigg) \epsilon^2
   \nn\\
   +&\bigg(-\frac{67}{9} \zeta _2+31 \zeta _4-\frac{2428}{81}\bigg) \epsilon +14 \zeta
   _3-\frac{404}{27}\bigg)
      \nn\\
   +&
  \CS{C_r N_f }\bigg(\bigg(\frac{56 }{27}\zeta _2+\frac{20 }{27}\zeta_3
 +\frac{5 }{12}\zeta _4+\frac{1952}{243}\bigg) \epsilon ^2+\bigg(\frac{10 }{9}\zeta_2+\frac{328}{81}\bigg) \epsilon +\frac{56}{27}\bigg)\bigg\}
   \nn\\
   +&\left(\frac{\als}{4 \pi}\right)^3\bigg\{
   \CS{C_A C_r N_f} 
   \bigg(\bigg(-\frac{8}{9} \zeta _3 \zeta _2+\frac{3925}{243} \zeta_2-\frac{1264 }{81}\zeta _3-\frac{172 }{3}\zeta _4-\frac{56 }{3}\zeta_5
   \nn\\
   +&\frac{716509}{4374}\bigg) \epsilon -\frac{412 }{81}\zeta _2-\frac{452 }{27}\zeta_3+\frac{10 }{3}\zeta _4+\frac{31313}{729}\bigg)
   \nn\\
   +&\CS{C_A^2 C_r}
   \bigg(\bigg(-\frac{680}{3} \zeta _3^2
   +\frac{440}{9} \zeta _2 \zeta _3+\frac{41548}{81} \zeta _3-\frac{8237 }{486}\zeta _2+\frac{1675 }{3}\zeta _4
   \nn\\
   +&308 \zeta_5-\frac{4409}{9} \zeta _6
   -\frac{7135981}{8748}\bigg) \epsilon
    -\frac{88}{3} \zeta_3 \zeta _2+\frac{3196 }{81}\zeta _2+\frac{6164 }{27}\zeta _3
    \nn\\
    +&\frac{77 }{3}\zeta_4-96 \zeta _5-\frac{297029}{1458}\bigg)
   \nn\\
   +&\CS{C_r N_f^2}
    \bigg(\bigg(-\frac{100 }{81}\zeta _2-\frac{160 }{27}\zeta _3-\frac{8 }{3}\zeta _4-\frac{11584}{2187}\bigg)
   \epsilon -\frac{16 }{9}\zeta _3-\frac{928}{729}\bigg)
   \nn\\
   +&\CS{C_F C_r N_f }\bigg(\bigg(-8 \zeta_3 \zeta _2+\frac{55 }{6}\zeta _2-\frac{1384 }{27}\zeta _3
   -\frac{76 }{3}\zeta _4-\frac{112 }{3}\zeta _5+\frac{42727}{324}\bigg) \epsilon 
   \nn\\
   -&\frac{152 }{9}\zeta_3-8 \zeta _4+\frac{1711}{54}\bigg)
   \bigg\}+\Ord(\als^4)\,.
\end{align}
The $\cO(\epsilon^0)$ terms at $\alpha_s^3$ appeared in \cite{Li:2016ctv}, the higher order terms in $\epsilon$ are new.

Using this result, we are now able to extract the rapidity anomalous dimension in QCD using the relation~\cite{Vladimirov:2016dll,Vladimirov:2017ksc}
\begin{align}
\gamma^{R}(\e^*)=\gamma^{S}\,,
\end{align}
where again $\e^*$ is known as the critical number of space-time dimension, whose value is determined by equation $\beta^*(\e^*)\equiv\frac{1}{\als}\beta(\als)-2\e^*=0$ order-by-order in QCD. Explicitly,
\begin{equation}
\e^* = - a_s \beta_0  - a_s^2 \beta_1 - a_s^3 \beta_2 - \cdots \,.
\end{equation}
The relation helps to build a bridge between different regimes of physics, such that the soft anomalous dimension can be obtained from rapidity anomalous dimension by substituting the dimensional regulator at its critical point, namely the QCD $\beta$-function (perturbative results for the $\beta$-function are collected in \App{sec:beta}). This allows us to derive an N$^4$LO relation between the soft and rapidity anomalous dimensions
\begin{align}
   \gamma^S_3-\gamma^R_3=&
   \CS{C_A C_r N_f}
   \beta_{0} \bigg(\frac{8}{9} \zeta _3 \zeta _2-\frac{3925}{243} \zeta _2
   +\frac{1264 }{81}\zeta_3+\frac{172 }{3}\zeta _4+\frac{56 }{3}\zeta _5-\frac{716509}{4374}\bigg)
   \nn\\
    +&\CS{C_A^2 C_r}
    \beta_{0} \bigg(\frac{680 }{3}\zeta _3^2-\frac{440}{9} \zeta _2 \zeta_3
    -\frac{41548 }{81}\zeta _3+\frac{8237 }{486}\zeta _2-\frac{1675 }{3}\zeta _4
    \nn\\
       -&308\zeta _5+\frac{4409 }{9}\zeta _6+\frac{7135981}{8748}\bigg) 
    \nn\\
    +&\CS{C_A C_r}
    \bigg(\beta_{0}^2 \bigg(\frac{46}{3} \zeta _3 \zeta _2
    -\frac{404 }{27}\zeta _2-\frac{134 }{27}\zeta_3-\frac{55}{24} \zeta _4+62 \zeta _5-\frac{14576}{243}\bigg)
    \nn\\
    +&\beta_{1}
   \bigg(\frac{67 }{9}\zeta _2-31 \zeta _4+\frac{2428}{81}\bigg)\bigg) 
   \nn\\
   +&\CS{C_r N_f}
   \bigg(\beta_{0}^2 \bigg(\frac{56 }{27}\zeta _2+\frac{20 }{27}\zeta _3+\frac{5 }{12}\zeta_4
   +\frac{1952}{243}\bigg)+\beta_{1} \bigg(-\frac{10 }{9}\zeta_2-\frac{328}{81}\bigg)\bigg) 
   \nn\\
   +&\CS{C_r N_f^2}
   \beta_{0} \bigg(\frac{100 }{81}\zeta_2+\frac{160}{27} \zeta _3+\frac{8 }{3}\zeta _4+\frac{11584}{2187}\bigg) 
   \nn\\
   +&\CS{C_F C_r N_f}
   \beta_{0} \bigg(8 \zeta _3 \zeta _2-\frac{55 }{6}\zeta _2+\frac{1384 }{27}\zeta_3
   +\frac{76}{3} \zeta _4+\frac{112 }{3}\zeta _5-\frac{42727}{324}\bigg) 
   \nn\\
   + &\CS{C_r}
   \bigg(\frac{9}{8} \beta_{0}^3 \zeta_4-\frac{4}{3} \beta_{1} \beta_{0} \zeta_3+\beta_{2} \zeta _2\bigg)
   \nn\\
   =&      %
   \CS{C_F^2 C_r N_f} \zeta _2 
   +\CS{C_r N_f^3}
  \bigg(\frac{8 }{81}\zeta_2-\frac{880 }{243}\zeta _3-\frac{52 }{27}\zeta _4+\frac{256}{6561}\bigg) 
   \nn\\
   +  &\CS{C_A C_r N_f^2}
   \bigg(\frac{56}{9} \zeta _3 \zeta _2+\frac{5353 }{1458}\zeta _2+\frac{616 }{243}\zeta_3
   -26 \zeta _4+\frac{136 }{9}\zeta _5+\frac{166639}{2187}\bigg) 
   \nn\\
   +&\CS{C_A C_F C_r N_f}
   \bigg(\frac{88}{3} \zeta _3 \zeta _2-\frac{539 }{9}\zeta _2+\frac{16016 }{81}\zeta _3
   +\frac{1394 }{9}\zeta _4+\frac{1232 }{9}\zeta _5-\frac{528269}{972}\bigg) 
   \nn\\
   + &\CS{C_A^3 C_r}
   \bigg(\frac{7480 }{9}\zeta _3^2+\frac{242}{9} \zeta _2 \zeta_3-\frac{486706}{243} \zeta _3
   -\frac{1273 }{729}\zeta _2-\frac{128191 }{54}\zeta_4
   \nn\\
   -&\frac{2662 }{9}\zeta _5+\frac{48499}{27}\zeta_6+\frac{66247055}{26244}\bigg)
   \nn\\
   +&\CS{C_F C_r N_f^2}
\bigg(-\frac{16}{3} \zeta _3 \zeta _2+\frac{86 }{9}\zeta _2-\frac{2912 }{81}\zeta _3-\frac{152 }{9}\zeta _4
-\frac{224 }{9}\zeta _5+\frac{46663}{486}\bigg) 
\nn\\
+&
\CS{C_A^2 C_r N_f}
   \bigg(-\frac{1360}{9} \zeta _3^2-\frac{352}{9} \zeta _2 \zeta _3+\frac{111724 }{243}\zeta_3
   -\frac{48257 }{1458}\zeta _2+\frac{2017 }{3}\zeta _4
   \nn\\
  -&\frac{88 }{3}\zeta_5 -\frac{8818 }{27}\zeta _6-\frac{648094}{729}\bigg) \,,
   \end{align}%
and this is the primary new perturbative ingredient presented in this paper.

Using the perturbative data from the splitting function and form factor anomalous dimensions presented earlier, this relation allows us to derive the four loop anomalous dimension in QCD
\begin{align}\label{eq:3_rap}
 \gamma^R_3 =&
\CS{\frac{ d_A^{abcd} d_r^{abcd} }{N_r}}
 \bigg(b(d^{(4)}_{FA})+\frac{1672 \zeta _3^2}{3}+\bigg(184 \zeta _4+\frac{3904}{9}\bigg) \zeta _3-\frac{1738}{9}\zeta _6
 \nn\\
 -&\frac{56}{3}\zeta _4+\frac{920}{9}\zeta _5+\zeta _2 \bigg(896 \zeta _3-512 \zeta _5+\frac{1088}{3}\bigg)-1742 \zeta _7-96\bigg)
 +\CS{C_A^3 C_r }
 \bigg(-\frac{b(d^{(4)}_{FA})}{24}
 \nn\\
 -&\frac{28290079}{8748}-\frac{5291 }{9}\zeta _3^2+\bigg(\frac{1057}{3} \zeta _4+\frac{300436}{81}\bigg) \zeta _3-\frac{45481}{27}\zeta _5-\frac{15895}{27}\zeta _6
 \nn\\
 +&\frac{2072}{9}\zeta _4+\frac{11071}{12}\zeta _7+\bigg(-\frac{6526}{9}\zeta _3+\frac{688}{3}\zeta _5+\frac{389083}{486}\bigg) \zeta _2\bigg)
 \nn\\
 +&\CS{N_f  C_A^2 C_r }
 \bigg(-\frac{1}{2} b\bigg(N_f C_F C_A C_r\bigg)-\frac{1}{4} b\bigg(N_f C_F^2 C_r\bigg)-\frac{b(d^{(4)}_{FF})}{48}-\frac{2146 \zeta _3^2}{9}-\frac{61913}{81}\zeta _3
 \nn\\
 -&\frac{4484}{27}\zeta _5+\frac{791}{54}\zeta _6+\frac{10906}{27}\zeta _4+\bigg(\frac{520}{3}\zeta _3-\frac{91067}{486}\bigg) \zeta _2+\frac{10761379}{11664}\bigg)
 \nn\\
 +&\CS{N_f  C_A C_F C_r}
  \bigg(b\bigg( N_f C_F C_A C_r\bigg)+\frac{1700 \zeta _3^2}{3}-\frac{30554}{27}\zeta _4-\frac{473}{9}\zeta _3+\frac{5476}{9}\zeta _5-\frac{2216}{9}\zeta _3 \zeta _2
  \nn\\
  +&\frac{2561}{54} \zeta _2-359 \zeta _6+\frac{2149049}{1944}\bigg)
 +\CS{N_f C_F^2 C_r }
 \bigg(b\bigg(C_F^3 N_f\bigg)-184 \zeta _3^2-\frac{1936}{3}\zeta _5+\frac{560}{9}\zeta _3
 \nn\\
 +&\frac{7334}{9}\zeta _6+\bigg(\frac{256}{3}\zeta _3-162\bigg) \zeta _2+167 \zeta _4-\frac{27949}{216}\bigg)
 +\CS{ N_f^2 C_A C_r}
 \bigg(-\frac{5564}{81}\zeta _3+\frac{40}{9}\zeta _4
 \nn\\
 +&\frac{368}{9}\zeta _5+\bigg(\frac{56}{9}\zeta _3+\frac{1688}{243}\bigg) \zeta _2-\frac{898033}{11664}\bigg) 
 + \CS{ N_f^2 C_F C_r}\bigg(\frac{40}{3}\zeta _4+\frac{1732}{27}\zeta _3+8 \zeta _5-\frac{110059}{972}\bigg)
 \nn\\
 +&\CS{N_f \frac{d_F^{ abcd} d_r^{ abcd} }{N_r}}
 \bigg(b(d^{(4)}_{FF})-\frac{608 }{3}\zeta _3^2-\frac{592}{9}\zeta _6+\frac{400}{3}\zeta _4+\frac{2656}{9}\zeta _3+\frac{10880}{9}\zeta _5-64 \zeta _3\zeta _2
 \nn\\
 -&\frac{2272}{3}\zeta _2+192\bigg)
 +\CS{C_r N_f^3}\bigg(-\frac{4}{9}\zeta _4+\frac{40}{9}\zeta _3+\frac{2608}{2187}\bigg) \,.
 \end{align}
For fundamental Wilson lines with generic $N_f$, the numeric results through to four loops are
\begin{align}
\gamma_R^{\rm q} =&\ a_s^2 (7.46333 + 2.76543 N_f) + a_s^3 (70.068 + 77.1286 N_f - 4.54662 N_f^2)
\nn
\\
& + a_s^4 (-350.8 + 2428 N_f - 378.3 N_f^2 + 8.072 N_f^3) + \cO(a_s^5) \,,
\end{align}
while for adjoint Wilson lines it is given by
\begin{align}
\gamma_R^{\rm g} =&\ a_s^2 ( 16.7925 + 6.22222 N_f) + a_s^3 ( 157.653 +173.539 N_f -10.2299 N_f^2)
\nn
\\
& + a_s^4 (333.8 +5506 N_f -851.2 N_f^2 + 18.16 N_f^3 ) + \cO(a_s^5) \,,
\end{align}
These are the primary results of our paper.  

\begin{figure}
\begin{center}
\includegraphics[width=0.6\textwidth]{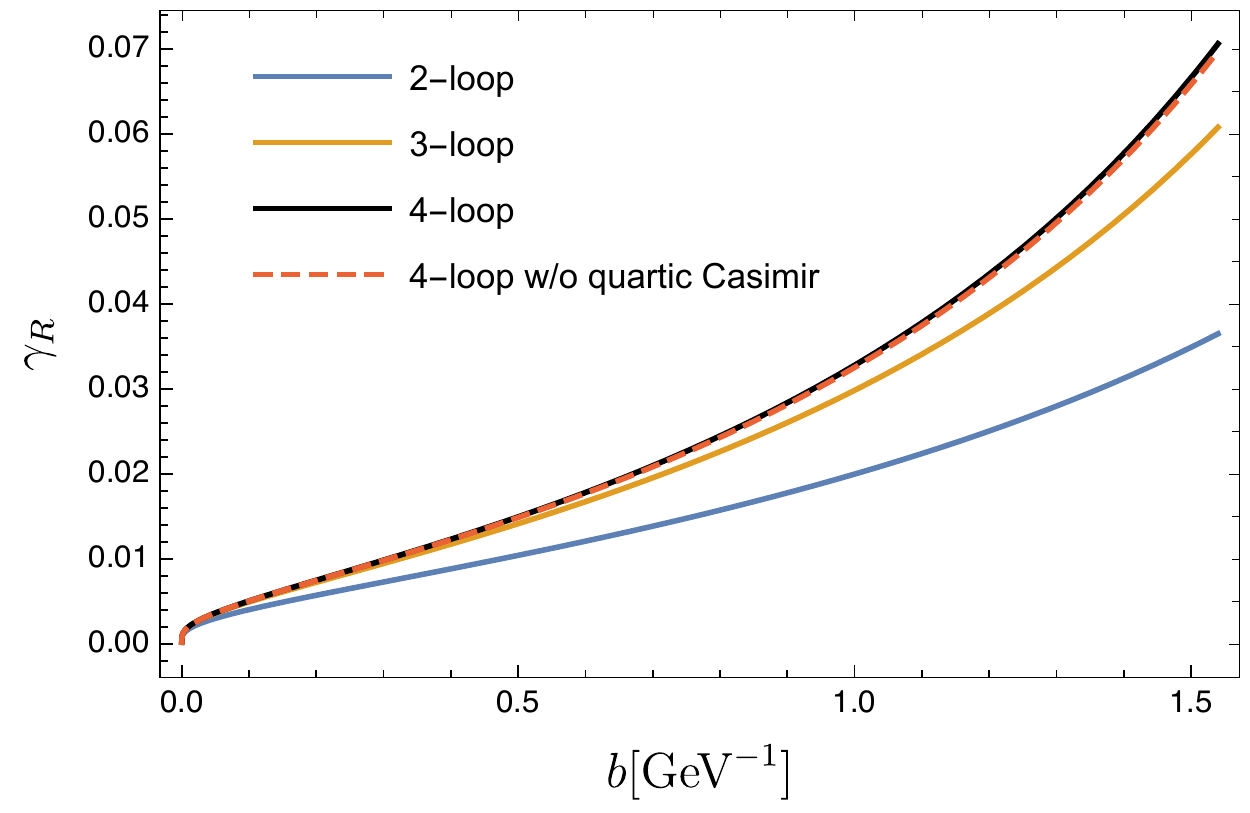}
\caption{In this plot we show the value of rapidity anomalous dimension for fundamental Wilson lines, with $N_f = 5$ and $\mu = b_0/b$, from two loops to four loops. We also show the four-loop results setting the quartic Casimir color factor to zero~(the dashed line). We note that the corrections from three loops to four loops are non-negligible, while the four-loop quartic Casimir corrections are very small.}
\label{fig:gammaR}
\end{center}
\end{figure}

In Fig.~\ref{fig:gammaR}, we plot the perturbative rapidity anomalous dimension through to four loops in QCD,\footnote{For non-perturbative calculations of the rapidity anomalous dimension or TMD soft function from lattice QCD, see e.g. \cite{Ji:2019ewn,Ebert:2018gzl,Ebert:2019tvc,Shanahan:2020zxr,LatticeParton:2020uhz,Li:2021wvl}.} using fundamental Wilson lines as an example. The four-loop corrections are non-negligible, and should be taken into account in future phenomenological studies. We also show the four-loop results without the quartic Casimir corrections, which illustrates that these corrections are small.

For the case of $q_T$ resummation rapidity renormalization group consistency implies that the rapidity anomalous dimension given in \Eq{eq:3_rap} also determines the beam function rapidity anomalous dimension to four loops.

\section{Conclusions}
\label{sec:conclusion}

In this paper we have computed the four loop rapidity anomalous dimension in QCD. This result was derived by exploiting a relation between the rapidity anomalous dimension and the soft anomalous dimension in a conformal theory, and computing the corrections to this relation arising from conformal symmetry breaking in QCD. At a technical level this was achieved by computing the fully differential soft function to $\cO(\epsilon)$ at three loops. We also described in some detail how this calculation was performed with the hope that it can be useful for performing calculations of other rapidity divergent soft functions.

In this paper we have also attempted to further clarify the nature of the rapidity anomalous dimension by presenting a simple proof relating the soft and rapidity anomalous dimensions using two equivalent representations of the the energy-energy correlator observable in the back-to-back limit. This relates the behavior of the four-point correlator \cite{Korchemsky:2019nzm}, with the rapidity anomalous dimension appearing in the time like factorization formula of \cite{Moult:2018jzp}. 

The four loop rapidity anomalous dimension enables the resummation of the EEC in the back-to-back region at N$^4$LL, which is the first event shape for which all ingredients are available to achieve this perturbative accuracy. It will be particularly interesting to study the convergence of the perturbative series, as well as the phenomenological implications for extractions of the strong coupling. For the case of $p_T$ resummation for the Higgs or electroweak $p_T$ spectra, the remaining ingredient required to achieve N$^4$LL resummation is the singlet four loop splitting functions. 

Our approach to computing the four loop rapidity anomalous dimension by exploiting its relation to other anomalous dimensions enabling it to be obtained through a three loop calculation illustrates the importance of better understanding relations between different anomalous dimensions in gauge theories. Much recent progress has been made along these lines \cite{Falcioni:2019nxk}. An interesting generalization of the case considered in this paper is to understand the relation of the rapidity anomalous dimension for observables with more than two colored lines, such as the transverse EEC as relevant for hadron colliders \cite{Gao:2019ojf}, and its relation to the soft anomalous dimension with multiple lines, which is known at three loops \cite{Almelid:2015jia, Almelid:2017qju}. In particular, it is interesting to understand whether the dipole conjecture holds for the rapidity anomalous dimension. Another rapidity type anomalous dimension that is difficult to compute directly is the Regge trajectory, which was recently computed in QCD to three loops \cite{Fadin:1995xg,Fadin:1995km,Fadin:1996tb,Korchemskaya:1996je,Blumlein:1998ib,DelDuca:2001gu,Bogdan:2002sr,Caola:2021izf,DelDuca:2021vjq,Falcioni:2021dgr,Falcioni:2021buo}. With the wealth of four loop data for anomalous dimensions, we believe it is important to understand its relation to other anomalous dimensions to enable its calculation to higher orders, and to obtain a more complete understanding of the relations between infrared anomalous dimensions.

\acknowledgments

We thank Alexey Vladimirov for useful discussions. We thank Claude Duhr, Bernhard Mistlberger and Gherardo Vita for coordinating the submission of their work. I.M. is supported by start-up funds from Yale University. H.X.Z is supported by National Natural Science Foundation of China under contract No. 11975200.

Note Added: 
  Simultaneously with this paper, the work of \cite{Duhr:2022yyp}  appeared, which also presented a result for the anomalous dimension, 
  we have verified that our result agrees.
\appendix

\section{Anomalous Dimensions and TMD Soft Function}
\label{sec:appendix}

In this Appendix, we collect the ingredients that entered into our calculations. We also summarize the results for the beam, soft and rapidity anomalous dimension at NNNNLO,
and the NNNLO TMD soft function to higher orders in the $\epsilon$ expansion.

\subsection{QCD Beta Function}
\label{sec:beta}

The QCD beta function is defined as
\begin{equation}
\frac{d\alpha_s}{d\ln\mu} = \beta(\alpha_s) = -2\alpha_s \sum_{n=0}^\infty \left( \frac{\alpha_s}{4 \pi} \right)^{n+1} \, \beta_n \, ,
\end{equation}
with~\cite{Baikov:2016tgj}
\begin{align}
\beta_0 &= \frac{11}{3} C_A - \frac{4}{3} T_F N_f \, ,
\\
\beta_1 &= \frac{34}{3} C_A^2 - \frac{20}{3} C_A T_F N_f - 4 C_F T_F N_f \, ,\nn
\\
\beta_2 &= \left(\frac{158 C_A}{27}+\frac{44 C_F}{9}\right) N_f^2 T_F^2 +\left(-\frac{205 C_A
   C_F}{9}-\frac{1415 C_A^2}{27}+2 C_F^2\right) N_f T_F  +\frac{2857 C_A^3}{54}\,. \nn
\end{align}

\subsection{Anomalous Dimensions}
\label{sec:AD}

We expand all anomalous dimensions in  $\alpha_s$ as
\begin{equation}
\gamma(\alpha_s) = \sum_{n=0}^\infty \left( \frac{\alpha_s}{4 \pi} \right)^{n+1} \, \gamma_n \, .
\end{equation}
The three-loop cusp, soft and rapidity anomalous dimension can be found in \cite{Moch:2004pa,Li:2014afw,Li:2016ctv,Vladimirov:2016dll}.  The four loop cusp anomalous dimension in QCD can be found in \cite{Henn:2019swt,vonManteuffel:2020vjv}. The numerical-analytical expressions for virtual and soft anomalous dimensions were obtained in~\cite{vonManteuffel:2020vjv,Das:2019btv,Agarwal:2021zft,Lee:2022nhh}. Based on these results, we obtain the full expressions for rapidity anomalous dimensions, albeit with some color coefficients a numerical value (see in Tab.~\ref{tab:Numeric_soft}).

The collected results for the coefficients up to $\Ord(\alpha_s^4)$ are given by 
\begin{align}
\Gcusp_{0} =& 4 \CS{C_r}\,, \nn
\\
\Gcusp_{1} =&  \left(\frac{268}{9}-8 \zeta_2\right) 
 \CS{C_A C_r} 
 -\frac{40 }{9} 
   \CS{C_r  N_f}\,, \nn
\\
\Gcusp_{2} =& \left(\frac{160 \zeta _2}{9}-\frac{112 \zeta _3}{3}-\frac{836}{27}\right)
 \CS{C_A C_r N_f }
 +\left(32 \zeta _3-\frac{110}{3}\right) 
 \CS{C_F C_r N_f }   \nn
   \\
 +&\left(-\frac{1072 \zeta_2}{9}+\frac{88 \zeta _3}{3}+88 \zeta _4+\frac{490}{3}\right)
    \CS{C_A^2 C_r}  
    -\frac{16}{27} 
   \CS{C_r N_f^2 }\,,\nn
  \\
\Gcusp_{3}= &
 \left(\frac{64}{27} \zeta _3-\frac{32}{81}\right)\CS{N_f^3 C_r}
+  \left(-\frac{224}{15} \zeta _2^2+\frac{2240}{27} \zeta _3-\frac{608}{81} \zeta _2+\frac{923}{81}\right) \CS{N_f^2 C_A C_r}\nn
\\
+& \CS{N_f^2 C_F C_r} \left(\frac{64}{5} \zeta _2^2-\frac{640}{9} \zeta _3+\frac{2392}{81}\right)
+ \bigg(
    \frac{2096}{9} \zeta _5
    +\frac{448}{3} \zeta _3 \zeta _2
    -\frac{352}{15} \zeta _2^2
    -\frac{23104}{27} \zeta _3\nn
    \\
    +&\frac{20320}{81} \zeta _2
    -\frac{24137}{81}
\bigg)\CS{N_f C_A^2 C_r} 
+ \bigg(
    160 \zeta _5
    -128 \zeta _3 \zeta _2
    -\frac{352}{5} \zeta _2^2 
    +\frac{3712}{9} \zeta _3
    +\frac{440}{3} \zeta _2\nn
    \\
   -&\frac{34066}{81}\bigg) \CS{N_f C_A C_F C_r}
+  \left(
    -320 \zeta _5
    +\frac{592}{3} \zeta _3
    +\frac{572}{9}
\right) \CS{N_f C_F^2 C_r}
+\bigg(
    -\frac{1280}{3} \zeta _5
    -\frac{256}{3} \zeta _3\nn
    \\
    +&256 \zeta _2
\bigg) \CS{N_f \frac{d^{abcd}_F d^{abcd}_r}{N_r}} 
+  \left(
    -384 \zeta _3^2
    -\frac{7936}{35} \zeta _2^3
    +\frac{3520}{3} \zeta _5
    +\frac{128}{3} \zeta _3
    -128 \zeta _2
\right)\CS{\frac{d^{abcd}_A d^{abcd}_r}{N_r}}\nn
\\
+& \left(
    -16 \zeta _3^2
    -\frac{20032}{105} \zeta _2^3
    -\frac{3608}{9} \zeta _5
    -\frac{352}{3} \zeta _3 \zeta _2
    +\frac{3608}{5} \zeta _2^2
    +\frac{20944}{27} \zeta _3
    -\frac{88400}{81} \zeta _2
    +\frac{84278}{81}
\right) \CS{C_A^3 C_r} \,.\nn
  \\
\gamma^S_0 =& 0 \, , \nn
\\
\gamma^S_1 =&  \left( -\frac{404}{27} + \frac{11\zeta_2}{3} + 14\zeta_3 \right)
 \CS{C_A  C_r}
 + \left( \frac{56}{27} - \frac{2\zeta_2}{3} \right)
 \CS{ C_r N_f }    \,, \nn
\\
\gamma^S_2  =
&\left(-\frac{88}{3} \zeta _3 \zeta _2
+\frac{6325 \zeta _2}{81}
+\frac{658 \zeta _3}{3}-88 \zeta_4-96 \zeta _5
-\frac{136781}{1458}\right) 
\CS{C_A^2 C_r}
+\bigg(\frac{20\zeta _2}{27}-\frac{56 \zeta _3}{27}\nn
\\
+&\frac{1040}{729}\bigg) 
\CS{C_r N_f^2 }
 + \left(-\frac{1414 \zeta _2}{81}-\frac{364 \zeta _3}{27}+24 \zeta _4
 +\frac{5921}{729}\right) 
   \CS{C_A C_r N_f }
   +\bigg(-2 \zeta _2-\frac{152 \zeta _3}{9}\nn
   \\
   -&8 \zeta_4
   +\frac{1711}{54}\bigg) 
   \CS{C_F C_r N_f }\,,\nn
   \\
        \gamma^S_3 =& 
\CS{\frac{ d_A^{abcd} d_r^{abcd} }{N_r}}
 \bigg(b(d^{(4)}_{FA})+\frac{1672 \zeta _3^2}{3}+\bigg(184 \zeta _4+\frac{3904}{9}\bigg) \zeta _3-\frac{1738}{9}\zeta _6
 \nn\\
 -&\frac{56}{3}\zeta _4+\frac{920}{9}\zeta _5+\zeta _2 \bigg(896 \zeta _3-512 \zeta _5+\frac{1088}{3}\bigg)-1742 \zeta _7-96\bigg)
 +\CS{C_A^3 C_r }
 \bigg(-\frac{b(d^{(4)}_{FA})}{24}
 \nn\\
 -&\frac{9311591}{13122}+\frac{2189 }{9}\zeta _3^2+\bigg(\frac{1057}{3} \zeta _4+\frac{414602}{243}\bigg) \zeta _3-\frac{53467}{27}\zeta _5+\frac{10868}{9}\zeta _6
 \nn\\
 -&\frac{115759}{54}\zeta _4+\frac{11071}{12}\zeta _7+\bigg(-\frac{6284}{9}\zeta _3+\frac{688}{3}\zeta _5+\frac{1164703}{1458}\bigg) \zeta _2\bigg)
 \nn\\
+&\CS{N_f  C_A^2 C_r} \bigg(-\frac{1}{2} b\bigg(C_A C_F^2 N_f\bigg)-\frac{1}{4} b\bigg(C_F^3 N_f\bigg)-\frac{b(d^{(4)}_{FF})}{48}-\frac{3506 \zeta _3^2}{9}-\frac{5615}{18}\zeta _6
\nn\\
-&\frac{74015}{243}\zeta _3-\frac{5276}{27}\zeta _5+\frac{29059}{27}\zeta _4+\bigg(\frac{1208}{9}\zeta _3-\frac{160729}{729}\bigg) \zeta _2+\frac{130625}{3888}\bigg)
\nn\\
+&\CS{N_f  C_A C_F C_r} \bigg(b\bigg(C_A C_F^2 N_f\bigg)+\frac{1700 \zeta _3^2}{3}-\frac{26372}{27}\zeta _4+\frac{11759}{81}\zeta _3+\frac{2236}{3}\zeta _5-\frac{1952}{9}\zeta _3 \zeta _2
\nn\\
-&\frac{673}{54} \zeta _2 -359 \zeta _6+\frac{1092511}{1944}\bigg)+ \CS{N_f  C_F^2 C_r} \bigg(b\bigg(C_F^3 N_f\bigg)-184 \zeta _3^2-\frac{1936}{3}\zeta _5+\frac{560}{9}\zeta _3
\nn\\
+&\frac{7334}{9}\zeta _6+\bigg(\frac{256}{3}\zeta _3-161\bigg) \zeta _2+167 \zeta _4-\frac{27949}{216}\bigg)+ \CS{C_A C_r N_f^2} \bigg(-\frac{16076}{243}\zeta _3-\frac{194}{9}\zeta _4
\nn\\
+&\frac{112}{9}\zeta _3 \zeta _2
+\frac{15481}{1458}\zeta _2+56 \zeta _5-\frac{27875}{34992}\bigg)
+\CS{C_F C_r N_f^2} \bigg(-\frac{152}{9}\zeta _5-\frac{32}{9}\zeta _4
+\frac{2284}{81}\zeta _3-\frac{16}{3}\zeta _3 \zeta _2
\nn\\
+&\frac{86}{9} \zeta _2
-\frac{16733}{972}\bigg)  +\CS{N_f \frac{d_F^{ abcd}d_r^{ abcd} }{N_r}}\bigg(b(d^{(4)}_{FF})-\frac{608 }{3}\zeta _3^2-\frac{592}{9}\zeta _6+\frac{400}{3}\zeta _4+\frac{2656}{9}\zeta _3
\nn\\
+&\frac{10880}{9}\zeta _5+\zeta _2 \bigg(-64 \zeta _3-\frac{2272}{3}\bigg)+192\bigg)+ \CS{C_r N_f^3}\bigg(-\frac{64}{27}\zeta _4+\frac{8}{81}\zeta _2
+\frac{200}{243}\zeta _3+\frac{8080}{6561}\bigg)\,.
   \nn
   \\
   \gamma^R_0 = &0 \, , \nn
\\
\gamma^R_1 = &  \left( -\frac{404}{27} + 14\zeta_3 \right) 
\CS{C_A  C_r }
+  \frac{56}{27} \CS{N_f  C_r } \,, \nn 
\\
\gamma^R_2 =&\left(-\frac{412 \zeta _2}{81}-\frac{452 \zeta _3}{27}
+\frac{10 \zeta_4}{3}+\frac{31313}{729}\right) 
\CS{C_A C_r N_f  }
   +\bigg(-\frac{88}{3} \zeta _3 \zeta_2+\frac{3196 \zeta _2}{81}
   +\frac{6164 \zeta _3}{27}\nn
   \\
   +&\frac{77 \zeta _4}{3}-96 \zeta_5 
   - \frac{297029}{1458}\bigg)\CS{C_A^2 C_r}
   + \left(-\frac{152 \zeta _3}{9}-8 \zeta_4
   +\frac{1711}{54}\right)  
   \CS{C_F C_r N_f  }
   +\bigg(-\frac{928}{729}\nn
   \\
   -&\frac{16 \zeta_3}{9}\bigg)
   \CS{ C_r N_f^2  }  \,,\nn
   \\
 \gamma^R_3 =&
\CS{\frac{ d_A^{abcd} d_r^{abcd} }{N_r}}
 \bigg(b(d^{(4)}_{FA})+\frac{1672 \zeta _3^2}{3}+\bigg(184 \zeta _4+\frac{3904}{9}\bigg) \zeta _3-\frac{1738}{9}\zeta _6
 \nn\\
 -&\frac{56}{3}\zeta _4+\frac{920}{9}\zeta _5+\zeta _2 \bigg(896 \zeta _3-512 \zeta _5+\frac{1088}{3}\bigg)-1742 \zeta _7-96\bigg)
 +\CS{C_A^3 C_r }
 \bigg(-\frac{b(d^{(4)}_{FA})}{24}
 \nn\\
 -&\frac{28290079}{8748}-\frac{5291 }{9}\zeta _3^2+\bigg(\frac{1057}{3} \zeta _4+\frac{300436}{81}\bigg) \zeta _3-\frac{45481}{27}\zeta _5-\frac{15895}{27}\zeta _6
 \nn\\
 +&\frac{2072}{9}\zeta _4+\frac{11071}{12}\zeta _7+\bigg(-\frac{6526}{9}\zeta _3+\frac{688}{3}\zeta _5+\frac{389083}{486}\bigg) \zeta _2\bigg)
 \nn\\
 +&\CS{N_f  C_A^2 C_r }
 \bigg(-\frac{1}{2} b\bigg(C_A C_F^2 N_f\bigg)-\frac{1}{4} b\bigg(C_F^3 N_f\bigg)-\frac{b(d^{(4)}_{FF})}{48}-\frac{2146 \zeta _3^2}{9}-\frac{61913}{81}\zeta _3
 \nn\\
 -&\frac{4484}{27}\zeta _5+\frac{791}{54}\zeta _6+\frac{10906}{27}\zeta _4+\bigg(\frac{520}{3}\zeta _3-\frac{91067}{486}\bigg) \zeta _2+\frac{10761379}{11664}\bigg)
 \nn\\
 +&\CS{N_f  C_A C_F C_r}
  \bigg(b\bigg(C_A C_F^2 N_f\bigg)+\frac{1700 \zeta _3^2}{3}-\frac{30554}{27}\zeta _4-\frac{473}{9}\zeta _3+\frac{5476}{9}\zeta _5-\frac{2216}{9}\zeta _3 \zeta _2
  \nn\\
  +&\frac{2561}{54} \zeta _2-359 \zeta _6+\frac{2149049}{1944}\bigg)
 +\CS{N_f C_F^2 C_r }
 \bigg(b\bigg(C_F^3 N_f\bigg)-184 \zeta _3^2-\frac{1936}{3}\zeta _5+\frac{560}{9}\zeta _3
 \nn\\
 +&\frac{7334}{9}\zeta _6+\bigg(\frac{256}{3}\zeta _3-162\bigg) \zeta _2+167 \zeta _4-\frac{27949}{216}\bigg)
 +\CS{ N_f^2 C_A C_r}
 \bigg(-\frac{5564}{81}\zeta _3+\frac{40}{9}\zeta _4
 \nn\\
 +&\frac{368}{9}\zeta _5+\bigg(\frac{56}{9}\zeta _3+\frac{1688}{243}\bigg) \zeta _2-\frac{898033}{11664}\bigg) 
 + \CS{ N_f^2 C_F C_r}\bigg(\frac{40}{3}\zeta _4+\frac{1732}{27}\zeta _3+8 \zeta _5-\frac{110059}{972}\bigg)
 \nn\\
 +&\CS{N_f \frac{d_F^{ abcd} d_r^{ abcd} }{N_r}}
 \bigg(b(d^{(4)}_{FF})-\frac{608 }{3}\zeta _3^2-\frac{592}{9}\zeta _6+\frac{400}{3}\zeta _4+\frac{2656}{9}\zeta _3+\frac{10880}{9}\zeta _5-64 \zeta _3\zeta _2
 \nn\\
 -&\frac{2272}{3}\zeta _2+192\bigg)
 +\CS{C_r N_f^3}\bigg(-\frac{4}{9}\zeta _4+\frac{40}{9}\zeta _3+\frac{2608}{2187}\bigg) \,.
 \end{align}
The cusp and soft and  rapidity anomalous dimensions  satisfy generalized Casimir scaling,
so above we only show the results for quark anomalous dimensions,
the corresponding gluon anomalous dimensions can be trivially obtained by generalized Casimir scaling.

The quark virtual anomalous dimensions  are
 \begin{align}
B_0^{\,\rm q} =& 3\CS{C_F} \, , \nn
\\
B_1^{\,\rm q} =&    
\left( \frac{3}{2} - 12\zeta_2 + 24\zeta_3 \right)
 \CS{C_F^2} 
+ \left( \frac{17}{6} + \frac{44 \zeta_2}{3} 
- 12\zeta_3 \right)  
\CS{C_A C_F} 
+ \left( -\frac{1}{3} - \frac{8\zeta_2}{3} \right)
 \CS{C_F  N_f}    , \nn
\\
B_2^{\,\rm q} = &
  \left(-\frac{1336 \zeta _2}{27}+\frac{200 \zeta _3}{9}+2\zeta _4+20\right) 
  \CS{C_A C_F N_f }
  +\left(\frac{20 \zeta _2}{3}-\frac{136 \zeta_3}{3}
   +\frac{116 \zeta _4}{3}-23\right) 
   \CS{C_F^2  N_f }\nn 
\\   
  +&\left(16 \zeta _3 \zeta _2-\frac{410 \zeta _2}{3}
  +\frac{844 \zeta _3}{3}-\frac{494 \zeta _4}{3}
   +120 \zeta _5+\frac{151}{4}\right) 
   \CS{C_A C_F^2}
   +\left(\frac{80 \zeta _2}{27}
   -\frac{16 \zeta _3}{9}
   -\frac{17}{9}\right)
   \CS{C_F N_f^2} \nn
   \\ 
   +&\left(\frac{4496\zeta _2}{27}
   -\frac{1552 \zeta _3}{9}-5 \zeta _4
   +40 \zeta _5-\frac{1657}{36}\right) 
   \CS{C_A^2 C_F}
   +\bigg(\frac{29}{2}-32 \zeta _3 \zeta _2
   +18 \zeta _2 +68 \zeta _3 \nn
   \\
   +& 144 \zeta_4-240 \zeta _5 \bigg) 
   \CS{C_F^3}\,, \nn 
   \\
 B_3^{\,\rm q} =&
 b(d^{(4)}_{FA}) \CS{ \frac{d_A^{abcd} d_F^{abcd}}{N_c}}
  + \CS{C_A^3 C_F} \bigg(-\frac{b(d^{(4)}_{FA})}{24}-\frac{371201}{648}+528 \zeta _3^2
  \nn\\
  +&\bigg(8 \zeta _4-\frac{153670}{81}\bigg) \zeta _3-\frac{11194}{27}\zeta _4+\frac{6046}{9}\zeta _6+\frac{11372}{9}\zeta _5+\frac{472}{3}\zeta _3 \zeta _2+504 \zeta _5\zeta _2
  \nn\\
  +&\frac{4582}{3}\zeta _2-2870 \zeta _7\bigg)
  +  \CS{N_f C_A^2 C_F} \bigg(-\frac{1}{2} b\bigg(C_A C_F^2 N_f\bigg)-\frac{1}{4} b\bigg(C_F^3 N_f\bigg)  -\frac{b(d^{(4)}_{FF})}{48}
  \nn\\
-&\frac{16 \zeta _3^2}{3}-\frac{248}{3}\zeta _5-\frac{137}{9}\zeta _3+\frac{16186}{27}\zeta _4+\bigg(-\frac{584}{9}\zeta _3-\frac{85175}{162}\bigg) \zeta _2
  -144 \zeta _6+\frac{353}{3}\bigg)
    \nn\\
  +& \CS{N_f C_A C_F^2} b\bigg(C_A C_F^2 N_f\bigg)
  + \CS{N_f C_F^3} b\bigg(C_F^3 N_f\bigg)
  + \bigg(-\frac{320}{9}\zeta _3-\frac{88}{9}\zeta _5-\frac{80}{9}\zeta _4
  \nn\\
  +&\bigg(\frac{80}{3}\zeta _3+\frac{3170}{81}\bigg) \zeta _2-\frac{193}{54}\bigg) \CS{ N_f^2 C_A C_F}
  +\bigg(-\frac{2104}{27}\zeta _4+\frac{56}{27}\zeta _3+\frac{368}{9}\zeta _5
-\frac{160}{9}\zeta _3 \zeta _2
\nn\\
+&\frac{1244}{27} \zeta _2
  -\frac{188}{27}\bigg) \CS{ N_f^2 C_F^2}
  + \CS{C_A C_F^3}
   \bigg(-\frac{2085}{4}+3220 \zeta _3^2+\bigg(128 \zeta _4-3260\bigg) \zeta _3
  \nn\\
  +&\frac{79297}{18}\zeta _6+2167 \zeta _4-976 \zeta _5+\zeta _2 \bigg(-\frac{1988}{3}\zeta _3+2064 \zeta _5+1167\bigg)-10920 \zeta _7\bigg)
  \nn\\
  +& \CS{C_A^2 C_F^2} \bigg(\frac{29639}{36}-\frac{7102 \zeta _3^2}{3}+\bigg(\frac{129662}{27}-32 \zeta _4\bigg) \zeta _3+\frac{5354}{9}\zeta _5
  -\frac{60850}{27}\zeta _4
  \nn\\
  -&\frac{5497}{2}\zeta _6+\zeta _2 \bigg(\frac{2096}{9}\zeta _3-2104 \zeta _5-\frac{46771}{27}\bigg)+8610 \zeta _7\bigg)
  +b(d^{(4)}_{FF})  \CS{N_f \frac{d_F^{ abcd}d_F^{ abcd}}{N_c}}
  \nn\\
  +&\bigg(-\frac{32}{27}\zeta _4+\frac{32}{81}\zeta _2+\frac{304}{81}\zeta _3-\frac{131}{81}\bigg)  \CS{C_F N_f^3}
  +\bigg(-1152 \zeta _3^2+64 \zeta _4 \zeta _3+2004 \zeta _3
  \nn\\
  -&342 \zeta _4+\zeta _2 \bigg(-120 \zeta _3-384 \zeta _5-450\bigg)-2520 \zeta _5-2111 \zeta _6+5880 \zeta _7+\frac{4873}{24}\bigg)  \CS{C_F^4}
  \,.\nn
 \end{align}
The gluon virtual anomalous dimensions  are
 \begin{align}
B_0^{\,\rm g} =& \frac{11}{3} \CS{C_A} - \frac{2}{3} \CS{ N_f} \,, \nn
\\
B_1^{\,\rm g} =& \CS{C_A^2} \left( \frac{32}{3}+ 12 \zeta_3 \right) + \left(  -\frac{8}{3} C_A -  2 C_F \right) \CS{N_f }   \,, \nn
\\
B_2^{\,\rm g} =& \CS{C_A^3}\left(-80\zeta_5-16\zeta_3\zeta_2+\frac{55}{3}\zeta_4+\frac{536}{3}\zeta_3+\frac{8}{3}\zeta_2+\frac{79}{2}\right)
\nn\\
+&\CS{C_A^2 N_f }\left(-\frac{10}{3}\zeta_4-\frac{80}{3}\zeta_3-\frac{8}{3}\zeta_2-\frac{233}{18}\right)
+\frac{27}{18}\CS{C_A N_f^2 }
-\frac{241}{18}\CS{C_A C_F N_f }
\nn\\
+& \CS{C_F^2 N_f } +\frac{11}{9}\CS{C_F N_f^2 }\,,
\nn\\
B_3^{\,\rm g} =&
\CS{N_f \frac{d_A^{abcd} d_F^{abcd}}{N_a}}
 \bigg(b(d^{(4)}_{FF})-\frac{1520}{3}\zeta _5-\frac{1496}{9}\zeta _6+\frac{1016}{3}\zeta _4+\frac{1312}{3}\zeta _3
 \nn\\
 +&\zeta _2 \bigg(544 \zeta _3-\frac{2368}{3}\bigg)+\frac{1952}{9}\bigg)
+\CS{C_A^4} \bigg(-\frac{b(d^{(4)}_{FA})}{24}+\frac{682 \zeta _3^2}{3}+\bigg(168 \zeta _4+\frac{48088}{27}\bigg) \zeta _3
\nn\\
-&\frac{14617}{9}\zeta _5-\frac{19129}{54}\zeta _6+\frac{8965}{54}\zeta _4+\zeta _2 \bigg(-\frac{3902}{9}\zeta _3+80 \zeta _5+\frac{2098}{27}\bigg)+700 \zeta _7+\frac{50387}{486}\bigg)
\nn\\
+ &
\CS{N_f C_A^3 }
\bigg(-\frac{1}{2} b\bigg(C_A C_F^2 N_f\bigg)-\frac{1}{4} b\bigg(C_F^3 N_f\bigg)-\frac{b(d^{(4)}_{FF})}{48}-\frac{1268 \zeta _3^2}{3}-\frac{22714}{27}\zeta _3+\frac{1777}{54}\zeta _6
\nn\\
+&\frac{919}{9}\zeta _5+\frac{7789}{18}\zeta _4+\bigg(\frac{1874}{9}\zeta _3-\frac{6155}{54}\bigg) \zeta _2-\frac{8075}{108}\bigg)
+\CS{N_f C_A^2 C_F} 
\bigg(b\bigg(C_A C_F^2 N_f\bigg)+\frac{1928 }{3}\zeta _3^2
\nn\\
-&\frac{27269}{27}\zeta _4-\frac{2879}{9}\zeta _6
+\frac{8854}{27}\zeta _3+\frac{6712}{9}\zeta _5
+\bigg(-\frac{2744}{9}\zeta _3+\frac{4198}{27}\bigg) \zeta _2+\frac{23566}{243}\bigg)
\nn\\
+&\CS{N_f C_A C_F^2 }
\bigg(b\bigg(C_F^3 N_f\bigg)-224 \zeta _3^2+\frac{2948}{9}\zeta _3+\frac{6434}{9}\zeta _6+\bigg(\frac{256}{3}\zeta _3-162\bigg) \zeta _2+204 \zeta _4
\nn\\
-&912 \zeta _5-\frac{2723}{27}\bigg)
+23 \CS{N_f C_F^3}
+ \bigg(\frac{160}{9}\zeta _3+\frac{3910}{243}\bigg)\CS{ N_f^2 C_A C_F}+\bigg(-\frac{8}{9}\zeta _5+\frac{200}{27}\zeta _4+\frac{289}{27}\zeta _3
\nn\\
+&\bigg(-\frac{32}{9}\zeta _3+\frac{37}{27}\bigg) \zeta _2+\frac{1352}{81}\bigg) \CS{N_f^2 C_A^2}+\bigg(-\frac{176}{9}\zeta _3+\frac{338}{27}\bigg)\CS{N_f^2 C_F^2}
+ \frac{5 }{243}\CS{C_A N_f^3}
\nn\\
+&\frac{154 }{243}\CS{C_FN_f^3}
+\CS{\frac{d_A^{ abcd} d_A^{ abcd} }{N_a}}\bigg(b(d^{(4)}_{FA})-\frac{784}{3}\zeta _3-\frac{508}{3}\zeta _4+\frac{748}{9}\zeta _6+\frac{760}{3}\zeta _5
+\zeta _2 \bigg(\frac{1184}{3}-272 \zeta _3\bigg)
\nn\\
-&\frac{800}{9}\bigg)
+\bigg(\frac{512}{3}\zeta _3-\frac{704}{9}\bigg)\CS{ N_f^2 \frac{d_F^{ abcd}d_F^{ abcd}}{N_a}}\,.
\end{align}

\subsection{TMD Soft Function}
\label{sec:TMDsoft}

In addition to the finite terms that were obtained in Ref.~\cite{Li:2016ctv}, in this Appendix we report a result to higher order in dimensional regulator $\e$, using the exponential regulator~\cite{Li:2016axz}.
Thanks to the Non-Abelian exponentiation theorem~\cite{Gatheral:1983cz,Frenkel:1984pz}, the TMD soft function allows an exponential form 
\begin{align}
s_1=&
\CS{C_r} \bigg[\epsilon ^3 \bigg(L_b^2 \bigg(-\zeta _2 L_r-\frac{4 \zeta _3}{3}\bigg)
+L_b \bigg(-\frac{4}{3} \zeta _3 L_r-\frac{27 \zeta_4}{4}\bigg)-\frac{1}{6} L_b^4 L_r-\frac{1}{3} \zeta _2L_b^3
\nn\\
   +&\frac{L_b^5}{30}-\frac{9}{4} \zeta _4 L_r-\frac{8}{3} \zeta _2 \zeta _3-\frac{16\zeta _5}{5}\bigg)
   +\epsilon ^2 \bigg(L_b \bigg(-2 \zeta _2 L_r-\frac{8 \zeta_3}{3}\bigg)-\frac{2}{3} L_b^3 L_r
 \nn\\
  -&\zeta _2 L_b^2 +\frac{L_b^4}{6}-\frac{4}{3} \zeta
   _3 L_r-\frac{27 \zeta _4}{4}\bigg)+\epsilon  \bigg(-2 L_b^2 L_r-2 \zeta _2
   L_b+\frac{2 L_b^3}{3}-2 \zeta _2 L_r-\frac{8 \zeta _3}{3}\bigg)
   \nn\\
   -&4 L_b L_r+2 L_b^2-2\zeta _2\bigg]\,,
   \nn\\[5mm]
    s_2=&\CS{C_r N_f} \bigg[-\frac{4 L_b^3}{9}+\bigg(\frac{4 L_r}{3}-\frac{20}{9}\bigg)
   L_b^2+\bigg(\frac{40 L_r}{9}+\frac{8 \zeta _2}{3}-\frac{112}{27}\bigg) L_b+\frac{112
   L_r}{27}+\frac{10 \zeta _2}{3}
   \nn\\
   +&\frac{28 \zeta _3}{9}+\epsilon 
   \bigg(-\frac{L_b^4}{3}+\bigg(\frac{4 L_r}{3}-\frac{40}{27}\bigg) L_b^3+\bigg(\frac{40
   L_r}{9}+\frac{10 \zeta _2}{3}-\frac{112}{27}\bigg) L_b^2
   \nn\\
   +&\bigg(\frac{20 \zeta
   _2}{3}+L_r \bigg(\frac{4 \zeta _2}{3}+\frac{224}{27}\bigg)+8 \zeta
   _3-\frac{656}{81}\bigg) L_b+\frac{56 \zeta _2}{9}+L_r \bigg(\frac{20 \zeta
   _2}{9}+\frac{656}{81}\bigg)
   \nn\\
   +&\frac{220 \zeta _3}{27}+\frac{35 \zeta
   _4}{3}-\frac{1952}{243}\bigg)+\epsilon ^2 \bigg(-\frac{7 L_b^5}{45}+\bigg(\frac{7
   L_r}{9}-\frac{20}{27}\bigg) L_b^4+\bigg(\frac{80 L_r}{27}+\frac{22 \zeta _2}{9}
      \nn\\
   -&\frac{224}{81}\bigg) L_b^3+\bigg(\frac{20 \zeta _2}{3}+L_r \bigg(2 \zeta _2
+\frac{224}{27}\bigg)+\frac{80 \zeta _3}{9}-\frac{656}{81}\bigg)
   L_b^2+\bigg(\frac{112 \zeta _2}{9}
   \nn\\
   +&L_r \bigg(\frac{40 \zeta _2}{9}+\frac{8 \zeta
   _3}{9}+\frac{1312}{81}\bigg)+\frac{440 \zeta _3}{27}+\frac{167 \zeta
   _4}{6}-\frac{3904}{243}\bigg) L_b+\frac{328 \zeta _2}{27}
   \nn\\
   +&\frac{44}{9} \zeta _2 \zeta
   _3+\frac{1232 \zeta _3}{81}+L_r \bigg(\frac{112 \zeta _2}{27}+\frac{40 \zeta
   _3}{27}+\frac{5 \zeta _4}{6}+\frac{3904}{243}\bigg)+\frac{485 \zeta _4}{18}
   \nn\\
   +&\frac{284\zeta _5}{15}-\frac{11680}{729}\bigg)-\frac{328}{81}\bigg]
   +\CS{C_A C_r} \bigg[\frac{22L_b^3}{9}+\bigg(-\frac{22 L_r}{3}-4 \zeta _2+\frac{134}{9}\bigg) L_b^2
   \nn\\
   +&\bigg(-\frac{44
   \zeta _2}{3}+L_r \bigg(8 \zeta _2-\frac{268}{9}\bigg)-28 \zeta
   _3+\frac{808}{27}\bigg) L_b-\frac{67 \zeta _2}{3}-\frac{154 \zeta _3}{9}
   \nn\\
   +&L_r \bigg(28 \zeta _3-\frac{808}{27}\bigg)+10 \zeta _4+\epsilon  \bigg(\frac{11
   L_b^4}{6}+\bigg(-\frac{22 L_r}{3}-\frac{8 \zeta _2}{3}+\frac{268}{27}\bigg)
   L_b^3
   \nn\\
   +&\bigg(-\frac{55 \zeta _2}{3}+L_r \bigg(8 \zeta _2-\frac{268}{9}\bigg)-28 \zeta
   _3+\frac{808}{27}\bigg) L_b^2+\bigg(-\frac{134 \zeta _2}{3}-44 \zeta _3
   \nn\\
   +&L_r\bigg(-\frac{22 \zeta _2}{3}+56 \zeta _3-\frac{1616}{27}\bigg)+20 \zeta
   _4+\frac{4856}{81}\bigg) L_b-\frac{404 \zeta _2}{9}+\frac{86}{3} \zeta _2 \zeta_3
   \nn\\
   -&\frac{1474 \zeta _3}{27}-\frac{385 \zeta _4}{6}+L_r \bigg(-\frac{134 \zeta
   _2}{9}+62 \zeta _4-\frac{4856}{81}\bigg)-2 \zeta _5+\frac{14576}{243}\bigg)
   \nn\\
   +&\epsilon^2 \bigg(\frac{77 L_b^5}{90}+\bigg(-\frac{77 L_r}{18}-\frac{4 \zeta
   _2}{3}+\frac{134}{27}\bigg) L_b^4+\bigg(-\frac{121 \zeta _2}{9}+L_r \bigg(\frac{16
   \zeta _2}{3}-\frac{536}{27}\bigg)
   \nn\\
   -&\frac{56 \zeta _3}{3}+\frac{1616}{81}\bigg)
   L_b^3+\bigg(-\frac{134 \zeta _2}{3}-\frac{440 \zeta _3}{9}+L_r \bigg(-11 \zeta _2+56
   \zeta _3-\frac{1616}{27}\bigg)
   \nn\\
   +&20 \zeta _4+\frac{4856}{81}\bigg)
   L_b^2+\bigg(\frac{172}{3} \zeta _3 \zeta _2-\frac{808 \zeta _2}{9}-\frac{2948 \zeta
   _3}{27}-\frac{1837 \zeta _4}{12}
   \nn\\
   +&L_r \bigg(-\frac{268 \zeta _2}{9}-\frac{44 \zeta
   _3}{9}+124 \zeta _4-\frac{9712}{81}\bigg)-4 \zeta _5+\frac{29152}{243}\bigg)
   L_b+\frac{206 \zeta _3^2}{3}
   \nn\\
   -&\frac{2428 \zeta _2}{27}-\frac{242}{9} \zeta _2 \zeta_3-\frac{8888 \zeta _3}{81}
   -\frac{6499 \zeta _4}{36}-\frac{1562 \zeta _5}{15}+L_r
   \bigg(\frac{92}{3} \zeta _3 \zeta _2
   \nn\\
   -&\frac{808 \zeta _2}{27}-\frac{268 \zeta
   _3}{27}-\frac{55 \zeta _4}{12}+124 \zeta _5-\frac{29152}{243}\bigg)+\frac{2009 \zeta
   _6}{8}+\frac{87472}{729}\bigg)+\frac{2428}{81}\bigg]\,,
   \nn\\[5mm]
     s_3 =&\CS{C_r C_A^2} \bigg[\frac{121 L_b^4}{27}+\bigg(-\frac{484 L_r}{27}-\frac{88 \zeta
   _2}{9}+\frac{3560}{81}\bigg) L_b^3+\bigg(-\frac{340 \zeta _2}{3}+L_r \bigg(\frac{88
   \zeta _2}{3}
      \nn\\
   -&\frac{3560}{27}\bigg)-88 \zeta _3+44 \zeta _4+\frac{15503}{81}\bigg)
   L_b^2+\bigg(\frac{176}{3} \zeta _3 \zeta _2-\frac{27752 \zeta _2}{81}-\frac{15232
   \zeta _3}{27}
   \nn\\
   +&L_r \bigg(\frac{1072 \zeta _2}{9}+176 \zeta _3-88 \zeta
   _4-\frac{31006}{81}\bigg)+\frac{748 \zeta _4}{3}+192 \zeta
   _5+\frac{297029}{729}\bigg) L_b
   \nn\\
   +&\frac{928 \zeta _3^2}{9}-\frac{297481 \zeta
   _2}{729}+\frac{1100}{9} \zeta _2 \zeta _3-\frac{151132 \zeta _3}{243}+\frac{3649
   \zeta _4}{27}
   \nn\\
   +&L_r \bigg(-\frac{176}{3} \zeta _3 \zeta _2+\frac{6392 \zeta
   _2}{81}+\frac{12328 \zeta _3}{27}+\frac{154 \zeta _4}{3}-192 \zeta
   _5-\frac{297029}{729}\bigg)
   \nn\\
   +&\frac{1804 \zeta _5}{9}-\frac{3086 \zeta _6}{27}+\epsilon
    \bigg(\frac{242 L_b^5}{45}+\bigg(-\frac{242 L_r}{9}-\frac{110 \zeta
   _2}{9}+\frac{4297}{81}\bigg) L_b^4+\bigg(-\frac{4270 \zeta _2}{27}
   \nn\\
   +&L_r \bigg(\frac{440
   \zeta _2}{9}-\frac{17188}{81}\bigg)-\frac{1408 \zeta _3}{9}+44 \zeta
   _4+\frac{64285}{243}\bigg) L_b^3+\bigg(88 \zeta _3 \zeta _2-\frac{18604 \zeta
   _2}{27}
   \nn\\
   -&\frac{9068 \zeta _3}{9}+L_r \bigg(\frac{1366 \zeta _2}{9}+\frac{1408 \zeta
   _3}{3}-132 \zeta _4-\frac{64285}{81}\bigg)+\frac{1342 \zeta _4}{3}+288 \zeta _5
   \nn\\
   +&\frac{403861}{486}\bigg) L_b^2+\bigg(\frac{928 \zeta _3^2}{3}+\frac{5192}{9} \zeta
   _2 \zeta _3-\frac{186008 \zeta _3}{81}-\frac{377473 \zeta _2}{243}-\frac{586 \zeta
   _4}{9}
   \nn\\
   +&L_r \bigg(-176 \zeta _3 \zeta _2+\frac{944 \zeta _2}{9}+\frac{12328 \zeta
   _3}{9}+\frac{1826 \zeta _4}{3}-576 \zeta _5
   -\frac{403861}{243}\bigg)
   \nn\\
   +&\frac{1760 \zeta
   _5}{3}-\frac{3086 \zeta _6}{9}+\frac{7135981}{4374}\bigg) L_b+\frac{18524 \zeta
   _3^2}{27}-\frac{6837355 \zeta _2}{4374}+\frac{20792}{27} \zeta _2 \zeta_3
   \nn\\
   -&\frac{1708132 \zeta _3}{729}-\frac{2816}{9} \zeta _3 \zeta _4-\frac{96557 \zeta
   _4}{81}-\frac{1448}{9} \zeta _2 \zeta _5-\frac{54476 \zeta _5}{135}
   \nn\\
   +&L_r
   \bigg(-\frac{1360}{3} \zeta _3^2+\frac{880}{9} \zeta _2 \zeta _3+\frac{83096 \zeta
   _3}{81}-\frac{8237 \zeta _2}{243}+\frac{3350 \zeta _4}{3}+616 \zeta _5
   \nn\\
   -&\frac{8818\zeta _6}{9}-\frac{7135981}{4374}\bigg)+\frac{121627 \zeta _6}{54}+\frac{712 \zeta
   _7}{9}+\frac{40870951}{26244}\bigg)+\frac{5211949}{13122}\bigg]
   \nn\\
   +&\CS{C_r C_A N_f}
   \bigg[-\frac{44 L_b^4}{27}+\bigg(\frac{176 L_r}{27}+\frac{16 \zeta
   _2}{9}-\frac{1156}{81}\bigg) L_b^3+\bigg(L_r \bigg(\frac{1156}{27}-\frac{16 \zeta
   _2}{3}\bigg)
   \nn\\
   +&\frac{256 \zeta _2}{9}-\frac{4102}{81}\bigg) L_b^2+\bigg(L_r
   \bigg(\frac{8204}{81}-\frac{160 \zeta _2}{9}\bigg)+\frac{7760 \zeta _2}{81}+\frac{1960
   \zeta _3}{27}-\frac{184 \zeta _4}{3}
   \nn\\
   -&\frac{62626}{729}\bigg) L_b+\frac{74530 \zeta
   _2}{729}+\frac{40}{9} \zeta _2 \zeta _3+\frac{8152 \zeta _3}{81}-\frac{416 \zeta
   _4}{27}+L_r \bigg(-\frac{824 \zeta _2}{81}
   \nn\\
   -&\frac{904 \zeta _3}{27}+\frac{20 \zeta
   _4}{3}+\frac{62626}{729}\bigg)-\frac{184 \zeta _5}{3}+\epsilon  \bigg(-\frac{88
   L_b^5}{45}+\bigg(\frac{88 L_r}{9}+\frac{20 \zeta _2}{9}-\frac{1400}{81}\bigg)
   L_b^4
   \nn\\
   +&\bigg(L_r \bigg(\frac{5600}{81}-\frac{80 \zeta _2}{9}\bigg)+\frac{1208 \zeta
   _2}{27}+\frac{112 \zeta _3}{9}-\frac{18002}{243}\bigg) L_b^3+\bigg(\frac{5434 \zeta
   _2}{27}
   \nn\\
   +&L_r \bigg(-\frac{152 \zeta _2}{9}-\frac{112 \zeta
   _3}{3}+\frac{18002}{81}\bigg)+\frac{1508 \zeta _3}{9}-\frac{316 \zeta
   _4}{3}-\frac{48241}{243}\bigg) L_b^2
   \nn\\
   +&\bigg(-\frac{224}{9} \zeta _3 \zeta
   _2+\frac{100162 \zeta _2}{243}+\frac{35912 \zeta _3}{81}+L_r \bigg(\frac{332 \zeta
   _2}{27}-\frac{904 \zeta _3}{9}-\frac{188 \zeta _4}{3}
   \nn\\
   +&\frac{96482}{243}\bigg)+\frac{1124 \zeta _4}{9}-\frac{544 \zeta
   _5}{3}-\frac{716509}{2187}\bigg) L_b-\frac{1160 \zeta _3^2}{27}+\frac{799511 \zeta
   _2}{2187}
   \nn\\
   -&\frac{452}{27} \zeta _2 \zeta _3+\frac{330472 \zeta _3}{729}+\frac{33430
   \zeta _4}{81}+L_r \bigg(-\frac{16}{9} \zeta _3 \zeta _2+\frac{7850 \zeta
   _2}{243}   -\frac{2528 \zeta _3}{81}
   \nn\\
 -&\frac{344 \zeta _4}{3}-\frac{112 \zeta
   _5}{3}+\frac{716509}{2187}\bigg)+\frac{15784 \zeta _5}{135}-\frac{10607 \zeta
   _6}{27}-\frac{375175}{1458}\bigg)-\frac{412765}{6561}\bigg] 
   \nn\\
   +&\CS{C_r N_f^2}
   \bigg[\frac{4 L_b^4}{27}+\bigg(\frac{80}{81}-\frac{16 L_r}{27}\bigg)
   L_b^3+\bigg(-\frac{80 L_r}{27}-\frac{16 \zeta _2}{9}+\frac{200}{81}\bigg)
   L_b^2+\bigg(-\frac{400 L_r}{81}
   \nn\\
   -&\frac{160 \zeta _2}{27}+\frac{1856}{729}\bigg)
   L_b-\frac{136 \zeta _2}{27}+L_r \bigg(-\frac{32 \zeta
   _3}{9}-\frac{1856}{729}\bigg)-\frac{560 \zeta _3}{243}-\frac{44 \zeta
   _4}{27}
   \nn\\
   +&\epsilon  \bigg(\frac{8 L_b^5}{45}+\bigg(\frac{100}{81}-\frac{8 L_r}{9}\bigg)
   L_b^4+\bigg(-\frac{400 L_r}{81}-\frac{88 \zeta _2}{27}+\frac{1048}{243}\bigg)
   L_b^3+\bigg(L_r \bigg(-\frac{8 \zeta _2}{9}
   \nn\\
   -&\frac{1048}{81}\bigg)-\frac{40 \zeta
   _2}{3}-\frac{16 \zeta _3}{3}+\frac{2240}{243}\bigg) L_b^2+\bigg(-\frac{632 \zeta
   _2}{27}+L_r \bigg(-\frac{80 \zeta _2}{27}-\frac{32 \zeta
   _3}{3}
   \nn\\
   -&\frac{4480}{243}\bigg)-\frac{160 \zeta _3}{9}-\frac{184 \zeta
   _4}{9}+\frac{23168}{2187}\bigg) L_b-\frac{12512 \zeta _2}{729}-\frac{32}{3} \zeta _2
   \zeta _3-\frac{3712 \zeta _3}{243}
   \nn\\
   +&L_r \bigg(-\frac{200 \zeta _2}{81}-\frac{320 \zeta
   _3}{27}-\frac{16 \zeta _4}{3}-\frac{23168}{2187}\bigg)-\frac{2540 \zeta
   _4}{81}-\frac{1744 \zeta
   _5}{135}+\frac{23168}{6561}\bigg)-\frac{256}{6561}\bigg]
   \nn\\
   +&\CS{C_F  C_r N_f}
    \bigg[-\frac{4L_b^3}{3}+\bigg(4 L_r+16 \zeta _3-\frac{55}{3}\bigg) L_b^2+\bigg(8 \zeta _2+L_r
   \bigg(\frac{110}{3}-32 \zeta _3\bigg)+\frac{304 \zeta _3}{9}
   \nn\\
   +&16 \zeta
   _4-\frac{1711}{27}\bigg) L_b+\frac{275 \zeta _2}{9}-\frac{80}{3} \zeta _2 \zeta
   _3+\frac{3488 \zeta _3}{81}+L_r \bigg(-\frac{304 \zeta _3}{9}-16 \zeta_4
   \nn\\
   +&\frac{1711}{27}\bigg)+\frac{152 \zeta _4}{9}+\frac{224 \zeta _5}{9}+\epsilon 
   \bigg(-\frac{4 L_b^4}{3}+\bigg(\frac{16 L_r}{3}+16 \zeta _3-\frac{55}{3}\bigg)
   L_b^3+\bigg(14 \zeta _2
   \nn\\
   +&L_r \bigg(55-48 \zeta _3\bigg)+\frac{152 \zeta _3}{3}+24 \zeta
   _4-\frac{1711}{18}\bigg) L_b^2+\bigg(-80 \zeta _3 \zeta _2+\frac{275 \zeta
   _2}{3}
   \nn\\
   +&\frac{3632 \zeta _3}{27}+L_r \bigg(4 \zeta _2-\frac{304 \zeta _3}{3}-48 \zeta
   _4+\frac{1711}{9}\bigg)+\frac{152 \zeta _4}{3}+\frac{224 \zeta
   _5}{3}-\frac{42727}{162}\bigg) L_b
   \nn\\
   -&\frac{736 \zeta _3^2}{9}+\frac{8555 \zeta
   _2}{54}-\frac{760}{9} \zeta _2 \zeta _3+\frac{47692 \zeta _3}{243}+\frac{2536 \zeta
   _4}{27}+L_r \bigg(-16 \zeta _3 \zeta _2
   \nn\\
   +&\frac{55 \zeta _2}{3}-\frac{2768 \zeta
   _3}{27}-\frac{152 \zeta _4}{3}-\frac{224 \zeta
   _5}{3}+\frac{42727}{162}\bigg)+\frac{2128 \zeta _5}{27}-\frac{50 \zeta
   _6}{3}-\frac{951775}{2916}\bigg)
   \nn\\
   -&\frac{42727}{486}\bigg]\,.
\end{align}

\bibliographystyle{JHEP}
\bibliography{AnomaNNNNLO}

\providecommand{\href}[2]{#2}\begingroup\raggedright\begin{thebibliography}{100}

\bibitem{Becher:2008cf}
T.~Becher and M.~D. Schwartz, {\it {A precise determination of $\alpha_s$ from
  LEP thrust data using effective field theory}},  {\em JHEP} {\bf 07} (2008)
  034, [\href{http://arxiv.org/abs/0803.0342}{{\tt arXiv:0803.0342}}].

\bibitem{Abbate:2010xh}
R.~Abbate, M.~Fickinger, A.~H. Hoang, V.~Mateu, and I.~W. Stewart, {\it {Thrust
  at $N^3LL$ with Power Corrections and a Precision Global Fit for
  alphas(mZ)}},  {\em Phys. Rev.} {\bf D83} (2011) 074021,
  [\href{http://arxiv.org/abs/1006.3080}{{\tt arXiv:1006.3080}}].

\bibitem{Hoang:2015hka}
A.~H. Hoang, D.~W. Kolodrubetz, V.~Mateu, and I.~W. Stewart, {\it {Precise
  determination of $\alpha_s$ from the $C$-parameter distribution}},  {\em
  Phys. Rev.} {\bf D91} (2015), no.~9 094018,
  [\href{http://arxiv.org/abs/1501.04111}{{\tt arXiv:1501.04111}}].

\bibitem{Kardos:2018kqj}
A.~Kardos, S.~Kluth, G.~Somogyi, Z.~Tulip{\'a}nt, and A.~Verbytskyi, {\it
  {Precise determination of $\alpha _{S}(M_Z)$ from a global fit of
  energy--energy correlation to NNLO+NNLL predictions}},  {\em Eur. Phys. J.}
  {\bf C78} (2018), no.~6 498, [\href{http://arxiv.org/abs/1804.09146}{{\tt
  arXiv:1804.09146}}].

\bibitem{Bishara:2016jga}
F.~Bishara, U.~Haisch, P.~F. Monni, and E.~Re, {\it {Constraining Light-Quark
  Yukawa Couplings from Higgs Distributions}},  {\em Phys. Rev. Lett.} {\bf
  118} (2017), no.~12 121801, [\href{http://arxiv.org/abs/1606.09253}{{\tt
  arXiv:1606.09253}}].

\bibitem{Soreq:2016rae}
Y.~Soreq, H.~X. Zhu, and J.~Zupan, {\it {Light quark Yukawa couplings from
  Higgs kinematics}},  {\em JHEP} {\bf 12} (2016) 045,
  [\href{http://arxiv.org/abs/1606.09621}{{\tt arXiv:1606.09621}}].

\bibitem{Bizon:2017rah}
W.~Bizon, P.~F. Monni, E.~Re, L.~Rottoli, and P.~Torrielli, {\it
  {Momentum-space resummation for transverse observables and the Higgs
  p$_{\perp}$ at N$^{3}$LL+NNLO}},  {\em JHEP} {\bf 02} (2018) 108,
  [\href{http://arxiv.org/abs/1705.09127}{{\tt arXiv:1705.09127}}].

\bibitem{Chen:2018pzu}
X.~Chen, T.~Gehrmann, E.~W.~N. Glover, A.~Huss, Y.~Li, D.~Neill, M.~Schulze,
  I.~W. Stewart, and H.~X. Zhu, {\it {Precise QCD Description of the Higgs
  Boson Transverse Momentum Spectrum}},  {\em Phys. Lett.} {\bf B788} (2019)
  425--430, [\href{http://arxiv.org/abs/1805.00736}{{\tt arXiv:1805.00736}}].

\bibitem{Bizon:2018foh}
W.~Bizo\'n, X.~Chen, A.~Gehrmann-De~Ridder, T.~Gehrmann, N.~Glover, A.~Huss,
  P.~F. Monni, E.~Re, L.~Rottoli, and P.~Torrielli, {\it {Fiducial
  distributions in Higgs and Drell-Yan production at N$^{3}$LL+NNLO}},  {\em
  JHEP} {\bf 12} (2018) 132, [\href{http://arxiv.org/abs/1805.05916}{{\tt
  arXiv:1805.05916}}].

\bibitem{Cieri:2018oms}
L.~Cieri, X.~Chen, T.~Gehrmann, E.~W.~N. Glover, and A.~Huss, {\it {Higgs boson
  production at the LHC using the $q_T$ subtraction formalism at N$^3$LO QCD}},
   {\em JHEP} {\bf 02} (2019) 096, [\href{http://arxiv.org/abs/1807.11501}{{\tt
  arXiv:1807.11501}}].

\bibitem{Bizon:2019zgf}
W.~Bizon, A.~Gehrmann-De~Ridder, T.~Gehrmann, N.~Glover, A.~Huss, P.~F. Monni,
  E.~Re, L.~Rottoli, and D.~M. Walker, {\it {The transverse momentum spectrum
  of weak gauge bosons at N ${}^3$ LL + NNLO}},  {\em Eur. Phys. J.} {\bf C79}
  (2019), no.~10 868, [\href{http://arxiv.org/abs/1905.05171}{{\tt
  arXiv:1905.05171}}].

\bibitem{Bacchetta:2019sam}
A.~Bacchetta, V.~Bertone, C.~Bissolotti, G.~Bozzi, F.~Delcarro, F.~Piacenza,
  and M.~Radici, {\it {Transverse-momentum-dependent parton distributions up to
  N$^{3}$LL from Drell-Yan data}},  {\em JHEP} {\bf 07} (2020) 117,
  [\href{http://arxiv.org/abs/1912.07550}{{\tt arXiv:1912.07550}}].

\bibitem{Kardos:2020gty}
A.~Kardos, A.~J. Larkoski, and Z.~trocsanyi, {\it {Groomed jet mass at high
  precision}},  \href{http://arxiv.org/abs/2002.00942}{{\tt arXiv:2002.00942}}.

\bibitem{Ebert:2020dfc}
M.~A. Ebert, J.~K.~L. Michel, I.~W. Stewart, and F.~J. Tackmann, {\it
  {Drell-Yan $q_{T}$ resummation of fiducial power corrections at N$^{3}$LL}},
  {\em JHEP} {\bf 04} (2021) 102, [\href{http://arxiv.org/abs/2006.11382}{{\tt
  arXiv:2006.11382}}].

\bibitem{Becher:2020ugp}
T.~Becher and T.~Neumann, {\it {Fiducial $q_T$ resummation of color-singlet
  processes at N$^3$LL+NNLO}},  {\em JHEP} {\bf 03} (2021) 199,
  [\href{http://arxiv.org/abs/2009.11437}{{\tt arXiv:2009.11437}}].

\bibitem{Billis:2021ecs}
G.~Billis, B.~Dehnadi, M.~A. Ebert, J.~K.~L. Michel, and F.~J. Tackmann, {\it
  {Higgs pT Spectrum and Total Cross Section with Fiducial Cuts at Third
  Resummed and Fixed Order in QCD}},  {\em Phys. Rev. Lett.} {\bf 127} (2021),
  no.~7 072001, [\href{http://arxiv.org/abs/2102.08039}{{\tt
  arXiv:2102.08039}}].

\bibitem{Camarda:2021ict}
S.~Camarda, L.~Cieri, and G.~Ferrera, {\it {Drell\textendash{}Yan lepton-pair
  production: qT resummation at N3LL accuracy and fiducial cross sections at
  N3LO}},  {\em Phys. Rev. D} {\bf 104} (2021), no.~11 L111503,
  [\href{http://arxiv.org/abs/2103.04974}{{\tt arXiv:2103.04974}}].

\bibitem{Re:2021con}
E.~Re, L.~Rottoli, and P.~Torrielli, {\it {Fiducial Higgs and Drell-Yan
  distributions at N$^3$LL$^\prime$+NNLO with RadISH}},
  \href{http://arxiv.org/abs/2104.07509}{{\tt arXiv:2104.07509}}.

\bibitem{Chen:2021vtu}
X.~Chen, T.~Gehrmann, N.~Glover, A.~Huss, T.-Z. Yang, and H.~X. Zhu, {\it
  {Dilepton Rapidity Distribution in Drell-Yan Production to Third Order in
  QCD}},  {\em Phys. Rev. Lett.} {\bf 128} (2022), no.~5 052001,
  [\href{http://arxiv.org/abs/2107.09085}{{\tt arXiv:2107.09085}}].

\bibitem{Chen:2022cgv}
X.~Chen, T.~Gehrmann, E.~W.~N. Glover, A.~Huss, P.~Monni, E.~Re, L.~Rottoli,
  and P.~Torrielli, {\it {Third order fiducial predictions for Drell-Yan at the
  LHC}},  \href{http://arxiv.org/abs/2203.01565}{{\tt arXiv:2203.01565}}.

\bibitem{Ju:2021lah}
W.-L. Ju and M.~Sch\"onherr, {\it {The q$_{T}$ and
  \ensuremath{\Delta}\ensuremath{\phi} spectra in W and Z production at the LHC
  at N$^{3}$LL'+N$^{2}$LO}},  {\em JHEP} {\bf 10} (2021) 088,
  [\href{http://arxiv.org/abs/2106.11260}{{\tt arXiv:2106.11260}}].

\bibitem{Baikov:2016tgj}
P.~A. Baikov, K.~G. Chetyrkin, and J.~H. K{\"u}hn, {\it {Five-Loop Running of
  the QCD coupling constant}},  {\em Phys. Rev. Lett.} {\bf 118} (2017), no.~8
  082002, [\href{http://arxiv.org/abs/1606.08659}{{\tt arXiv:1606.08659}}].

\bibitem{Herzog:2017ohr}
F.~Herzog, B.~Ruijl, T.~Ueda, J.~Vermaseren, and A.~Vogt, {\it {The five-loop
  beta function of Yang-Mills theory with fermions}},  {\em JHEP} {\bf 02}
  (2017) 090, [\href{http://arxiv.org/abs/1701.01404}{{\tt arXiv:1701.01404}}].

\bibitem{Luthe:2017ttg}
T.~Luthe, A.~Maier, P.~Marquard, and Y.~Schroder, {\it {The five-loop Beta
  function for a general gauge group and anomalous dimensions beyond Feynman
  gauge}},  {\em JHEP} {\bf 10} (2017) 166,
  [\href{http://arxiv.org/abs/1709.07718}{{\tt arXiv:1709.07718}}].

\bibitem{Chetyrkin:2017bjc}
K.~Chetyrkin, G.~Falcioni, F.~Herzog, and J.~Vermaseren, {\it {Five-loop
  renormalisation of QCD in covariant gauges}},  {\em JHEP} {\bf 10} (2017)
  179, [\href{http://arxiv.org/abs/1709.08541}{{\tt arXiv:1709.08541}}].
  [Addendum: JHEP 12, 006 (2017)].

\bibitem{Ruijl:2016pkm}
B.~Ruijl, T.~Ueda, J.~A.~M. Vermaseren, J.~Davies, and A.~Vogt, {\it {First
  Forcer results on deep-inelastic scattering and related quantities}},  {\em
  PoS} {\bf LL2016} (2016) 071, [\href{http://arxiv.org/abs/1605.08408}{{\tt
  arXiv:1605.08408}}].

\bibitem{Moch:2017uml}
S.~Moch, B.~Ruijl, T.~Ueda, J.~A.~M. Vermaseren, and A.~Vogt, {\it {Four-Loop
  Non-Singlet Splitting Functions in the Planar Limit and Beyond}},  {\em JHEP}
  {\bf 10} (2017) 041, [\href{http://arxiv.org/abs/1707.08315}{{\tt
  arXiv:1707.08315}}].

\bibitem{Herzog:2018kwj}
F.~Herzog, S.~Moch, B.~Ruijl, T.~Ueda, J.~A.~M. Vermaseren, and A.~Vogt, {\it
  {Five-loop contributions to low-N non-singlet anomalous dimensions in QCD}},
  {\em Phys. Lett.} {\bf B790} (2019) 436--443,
  [\href{http://arxiv.org/abs/1812.11818}{{\tt arXiv:1812.11818}}].

\bibitem{Moch:2018wjh}
S.~Moch, B.~Ruijl, T.~Ueda, J.~A.~M. Vermaseren, and A.~Vogt, {\it {On quartic
  colour factors in splitting functions and the gluon cusp anomalous
  dimension}},  {\em Phys. Lett.} {\bf B782} (2018) 627--632,
  [\href{http://arxiv.org/abs/1805.09638}{{\tt arXiv:1805.09638}}].

\bibitem{Grozin:2015kna}
A.~Grozin, J.~M. Henn, G.~P. Korchemsky, and P.~Marquard, {\it {The three-loop
  cusp anomalous dimension in QCD and its supersymmetric extensions}},  {\em
  JHEP} {\bf 01} (2016) 140, [\href{http://arxiv.org/abs/1510.07803}{{\tt
  arXiv:1510.07803}}].

\bibitem{Henn:2016men}
J.~M. Henn, A.~V. Smirnov, V.~A. Smirnov, and M.~Steinhauser, {\it {A planar
  four-loop form factor and cusp anomalous dimension in QCD}},  {\em JHEP} {\bf
  05} (2016) 066, [\href{http://arxiv.org/abs/1604.03126}{{\tt
  arXiv:1604.03126}}].

\bibitem{vonManteuffel:2016xki}
A.~von Manteuffel and R.~M. Schabinger, {\it {Quark and gluon form factors to
  four-loop order in QCD: the $N_f^3$ contributions}},  {\em Phys. Rev.} {\bf
  D95} (2017), no.~3 034030, [\href{http://arxiv.org/abs/1611.00795}{{\tt
  arXiv:1611.00795}}].

\bibitem{Lee:2016ixa}
J.~Henn, A.~V. Smirnov, V.~A. Smirnov, M.~Steinhauser, and R.~N. Lee, {\it
  {Four-loop photon quark form factor and cusp anomalous dimension in the
  large-$N_c$ limit of QCD}},  {\em JHEP} {\bf 03} (2017) 139,
  [\href{http://arxiv.org/abs/1612.04389}{{\tt arXiv:1612.04389}}].

\bibitem{Lee:2017mip}
R.~N. Lee, A.~V. Smirnov, V.~A. Smirnov, and M.~Steinhauser, {\it {The $n_f^2$
  contributions to fermionic four-loop form factors}},  {\em Phys. Rev.} {\bf
  D96} (2017), no.~1 014008, [\href{http://arxiv.org/abs/1705.06862}{{\tt
  arXiv:1705.06862}}].

\bibitem{Henn:2019rmi}
J.~M. Henn, T.~Peraro, M.~Stahlhofen, and P.~Wasser, {\it {Matter dependence of
  the four-loop cusp anomalous dimension}},  {\em Phys. Rev. Lett.} {\bf 122}
  (2019), no.~20 201602, [\href{http://arxiv.org/abs/1901.03693}{{\tt
  arXiv:1901.03693}}].

\bibitem{Lee:2019zop}
R.~N. Lee, A.~V. Smirnov, V.~A. Smirnov, and M.~Steinhauser, {\it {Four-loop
  quark form factor with quartic fundamental colour factor}},  {\em JHEP} {\bf
  02} (2019) 172, [\href{http://arxiv.org/abs/1901.02898}{{\tt
  arXiv:1901.02898}}].

\bibitem{vonManteuffel:2019wbj}
A.~von Manteuffel and R.~M. Schabinger, {\it {Quark and gluon form factors in
  four loop QCD: The $N_f^2$ and $N_{q\gamma} N_f$ contributions}},  {\em Phys.
  Rev.} {\bf D99} (2019), no.~9 094014,
  [\href{http://arxiv.org/abs/1902.08208}{{\tt arXiv:1902.08208}}].

\bibitem{Bruser:2019auj}
R.~Br{\"u}ser, A.~Grozin, J.~M. Henn, and M.~Stahlhofen, {\it {Matter
  dependence of the four-loop QCD cusp anomalous dimension: from small angles
  to all angles}},  {\em JHEP} {\bf 05} (2019) 186,
  [\href{http://arxiv.org/abs/1902.05076}{{\tt arXiv:1902.05076}}].

\bibitem{Henn:2019swt}
J.~M. Henn, G.~P. Korchemsky, and B.~Mistlberger, {\it {The full four-loop cusp
  anomalous dimension in $\mathcal{N}=4$ super Yang-Mills and QCD}},  {\em
  JHEP} {\bf 04} (2020) 018, [\href{http://arxiv.org/abs/1911.10174}{{\tt
  arXiv:1911.10174}}].

\bibitem{vonManteuffel:2020vjv}
A.~von Manteuffel, E.~Panzer, and R.~M. Schabinger, {\it {Analytic four-loop
  anomalous dimensions in massless QCD from form factors}},
  \href{http://arxiv.org/abs/2002.04617}{{\tt arXiv:2002.04617}}.

\bibitem{Das:2019btv}
G.~Das, S.-O. Moch, and A.~Vogt, {\it {Soft corrections to inclusive
  deep-inelastic scattering at four loops and beyond}},
  \href{http://arxiv.org/abs/1912.12920}{{\tt arXiv:1912.12920}}.

\bibitem{Agarwal:2021zft}
B.~Agarwal, A.~von Manteuffel, E.~Panzer, and R.~M. Schabinger, {\it {Four-loop
  collinear anomalous dimensions in QCD and N=4 super Yang-Mills}},  {\em Phys.
  Lett. B} {\bf 820} (2021) 136503,
  [\href{http://arxiv.org/abs/2102.09725}{{\tt arXiv:2102.09725}}].

\bibitem{Lee:2021uqq}
R.~N. Lee, A.~von Manteuffel, R.~M. Schabinger, A.~V. Smirnov, V.~A. Smirnov,
  and M.~Steinhauser, {\it {Fermionic corrections to quark and gluon form
  factors in four-loop QCD}},  {\em Phys. Rev. D} {\bf 104} (2021), no.~7
  074008, [\href{http://arxiv.org/abs/2105.11504}{{\tt arXiv:2105.11504}}].

\bibitem{Lee:2022nhh}
R.~N. Lee, A.~von Manteuffel, R.~M. Schabinger, A.~V. Smirnov, V.~A. Smirnov,
  and M.~Steinhauser, {\it {Quark and gluon form factors in four-loop QCD}},
  \href{http://arxiv.org/abs/2202.04660}{{\tt arXiv:2202.04660}}.

\bibitem{Chakraborty:2022yan}
A.~Chakraborty, T.~Huber, R.~N. Lee, A.~von Manteuffel, R.~M. Schabinger, A.~V.
  Smirnov, V.~A. Smirnov, and M.~Steinhauser, {\it {The $Hb\bar{b}$ vertex at
  four loops and hard matching coefficients in SCET for various currents}},
  \href{http://arxiv.org/abs/2204.02422}{{\tt arXiv:2204.02422}}.

\bibitem{Das:2020adl}
G.~Das, S.~Moch, and A.~Vogt, {\it {Approximate four-loop QCD corrections to
  the Higgs-boson production cross section}},
  \href{http://arxiv.org/abs/2004.00563}{{\tt arXiv:2004.00563}}.

\bibitem{Basham:1978bw}
C.~L. Basham, L.~S. Brown, S.~D. Ellis, and S.~T. Love, {\it {Energy
  Correlations in electron - Positron Annihilation: Testing QCD}},  {\em Phys.
  Rev. Lett.} {\bf 41} (1978) 1585.

\bibitem{Collins:1981uk}
J.~C. Collins and D.~E. Soper, {\it {Back-To-Back Jets in QCD}},  {\em Nucl.
  Phys.} {\bf B193} (1981) 381. [Erratum: Nucl. Phys.B213,545(1983)].

\bibitem{Collins:1981va}
J.~C. Collins and D.~E. Soper, {\it {Back-To-Back Jets: Fourier Transform from
  B to K-Transverse}},  {\em Nucl. Phys.} {\bf B197} (1982) 446--476.

\bibitem{Collins:1984kg}
J.~C. Collins, D.~E. Soper, and G.~F. Sterman, {\it {Transverse Momentum
  Distribution in Drell-Yan Pair and W and Z Boson Production}},  {\em Nucl.
  Phys.} {\bf B250} (1985) 199--224.

\bibitem{Chiu:2011qc}
J.-y. Chiu, A.~Jain, D.~Neill, and I.~Z. Rothstein, {\it {The Rapidity
  Renormalization Group}},  {\em Phys. Rev. Lett.} {\bf 108} (2012) 151601,
  [\href{http://arxiv.org/abs/1104.0881}{{\tt arXiv:1104.0881}}].

\bibitem{Chiu:2012ir}
J.-Y. Chiu, A.~Jain, D.~Neill, and I.~Z. Rothstein, {\it {A Formalism for the
  Systematic Treatment of Rapidity Logarithms in Quantum Field Theory}},  {\em
  JHEP} {\bf 05} (2012) 084, [\href{http://arxiv.org/abs/1202.0814}{{\tt
  arXiv:1202.0814}}].

\bibitem{Bauer:2000ew}
C.~W. Bauer, S.~Fleming, and M.~E. Luke, {\it {Summing Sudakov logarithms in $B
  \to X_s \gamma$ in effective field theory}},  {\em Phys. Rev.} {\bf D63}
  (2000) 014006, [\href{http://arxiv.org/abs/hep-ph/0005275}{{\tt
  hep-ph/0005275}}].

\bibitem{Bauer:2000yr}
C.~W. Bauer, S.~Fleming, D.~Pirjol, and I.~W. Stewart, {\it {An Effective field
  theory for collinear and soft gluons: Heavy to light decays}},  {\em Phys.
  Rev.} {\bf D63} (2001) 114020,
  [\href{http://arxiv.org/abs/hep-ph/0011336}{{\tt hep-ph/0011336}}].

\bibitem{Bauer:2001ct}
C.~W. Bauer and I.~W. Stewart, {\it {Invariant operators in collinear effective
  theory}},  {\em Phys. Lett.} {\bf B516} (2001) 134--142,
  [\href{http://arxiv.org/abs/hep-ph/0107001}{{\tt hep-ph/0107001}}].

\bibitem{Bauer:2001yt}
C.~W. Bauer, D.~Pirjol, and I.~W. Stewart, {\it {Soft collinear factorization
  in effective field theory}},  {\em Phys. Rev.} {\bf D65} (2002) 054022,
  [\href{http://arxiv.org/abs/hep-ph/0109045}{{\tt hep-ph/0109045}}].

\bibitem{Bauer:2002aj}
C.~W. Bauer, D.~Pirjol, and I.~W. Stewart, {\it {Factorization and endpoint
  singularities in heavy to light decays}},  {\em Phys. Rev.} {\bf D67} (2003)
  071502, [\href{http://arxiv.org/abs/hep-ph/0211069}{{\tt hep-ph/0211069}}].

\bibitem{Li:2016ctv}
Y.~Li and H.~X. Zhu, {\it {Bootstrapping Rapidity Anomalous Dimensions for
  Transverse-Momentum Resummation}},  {\em Phys. Rev. Lett.} {\bf 118} (2017),
  no.~2 022004, [\href{http://arxiv.org/abs/1604.01404}{{\tt
  arXiv:1604.01404}}].

\bibitem{Vladimirov:2016dll}
A.~A. Vladimirov, {\it {Correspondence between Soft and Rapidity Anomalous
  Dimensions}},  {\em Phys. Rev. Lett.} {\bf 118} (2017), no.~6 062001,
  [\href{http://arxiv.org/abs/1610.05791}{{\tt arXiv:1610.05791}}].

\bibitem{Vladimirov:2017ksc}
A.~Vladimirov, {\it {Structure of rapidity divergences in multi-parton
  scattering soft factors}},  {\em JHEP} {\bf 04} (2018) 045,
  [\href{http://arxiv.org/abs/1707.07606}{{\tt arXiv:1707.07606}}].

\bibitem{Li:2011zp}
Y.~Li, S.~Mantry, and F.~Petriello, {\it {An Exclusive Soft Function for
  Drell-Yan at Next-to-Next-to-Leading Order}},  {\em Phys. Rev.} {\bf D84}
  (2011) 094014, [\href{http://arxiv.org/abs/1105.5171}{{\tt
  arXiv:1105.5171}}].

\bibitem{moult_talk}
I.~Moult, ``World scet (2020),
  https://indico.cern.ch/event/911911/contributions/3872185/.''

\bibitem{zhu_talk}
H.~X. Zhu, ``Online talk at qcd online seminar, april 24 (2020).''

\bibitem{Korchemsky:2019nzm}
G.~P. Korchemsky, {\it {Energy correlations in the end-point region}},  {\em
  JHEP} {\bf 01} (2020) 008, [\href{http://arxiv.org/abs/1905.01444}{{\tt
  arXiv:1905.01444}}].

\bibitem{Moult:2018jzp}
I.~Moult and H.~X. Zhu, {\it {Simplicity from Recoil: The Three-Loop Soft
  Function and Factorization for the Energy-Energy Correlation}},  {\em JHEP}
  {\bf 08} (2018) 160, [\href{http://arxiv.org/abs/1801.02627}{{\tt
  arXiv:1801.02627}}].

\bibitem{Korchemsky:1988si}
G.~P. Korchemsky, {\it {Asymptotics of the Altarelli-Parisi-Lipatov Evolution
  Kernels of Parton Distributions}},  {\em Mod. Phys. Lett.} {\bf A4} (1989)
  1257--1276.

\bibitem{Alday:2007mf}
L.~F. Alday and J.~M. Maldacena, {\it {Comments on operators with large spin}},
   {\em JHEP} {\bf 11} (2007) 019, [\href{http://arxiv.org/abs/0708.0672}{{\tt
  arXiv:0708.0672}}].

\bibitem{Korchemsky:1992xv}
G.~P. Korchemsky and G.~Marchesini, {\it {Structure function for large x and
  renormalization of Wilson loop}},  {\em Nucl. Phys.} {\bf B406} (1993)
  225--258, [\href{http://arxiv.org/abs/hep-ph/9210281}{{\tt hep-ph/9210281}}].

\bibitem{Belitsky:2008mg}
A.~V. Belitsky, G.~P. Korchemsky, and R.~S. Pasechnik, {\it {Fine structure of
  anomalous dimensions in N=4 super Yang-Mills theory}},  {\em Nucl. Phys.}
  {\bf B809} (2009) 244--278, [\href{http://arxiv.org/abs/0806.3657}{{\tt
  arXiv:0806.3657}}].

\bibitem{Korchemsky:1987wg}
G.~Korchemsky and A.~Radyushkin, {\it {Renormalization of the Wilson Loops
  Beyond the Leading Order}},  {\em Nucl.Phys.} {\bf B283} (1987) 342--364.

\bibitem{Beisert:2006ez}
N.~Beisert, B.~Eden, and M.~Staudacher, {\it {Transcendentality and Crossing}},
   {\em J. Stat. Mech.} {\bf 0701} (2007) P01021,
  [\href{http://arxiv.org/abs/hep-th/0610251}{{\tt hep-th/0610251}}].

\bibitem{Eden:2006rx}
B.~Eden and M.~Staudacher, {\it {Integrability and transcendentality}},  {\em
  J. Stat. Mech.} {\bf 0611} (2006) P11014,
  [\href{http://arxiv.org/abs/hep-th/0603157}{{\tt hep-th/0603157}}].

\bibitem{Freyhult:2007pz}
L.~Freyhult, A.~Rej, and M.~Staudacher, {\it {A Generalized Scaling Function
  for AdS/CFT}},  {\em J. Stat. Mech.} {\bf 0807} (2008) P07015,
  [\href{http://arxiv.org/abs/0712.2743}{{\tt arXiv:0712.2743}}].

\bibitem{Freyhult:2009my}
L.~Freyhult and S.~Zieme, {\it {The virtual scaling function of AdS/CFT}},
  {\em Phys. Rev.} {\bf D79} (2009) 105009,
  [\href{http://arxiv.org/abs/0901.2749}{{\tt arXiv:0901.2749}}].

\bibitem{Fioravanti:2009xt}
D.~Fioravanti, P.~Grinza, and M.~Rossi, {\it {Beyond cusp anomalous dimension
  from integrability}},  {\em Phys. Lett.} {\bf B675} (2009) 137--144,
  [\href{http://arxiv.org/abs/0901.3161}{{\tt arXiv:0901.3161}}].

\bibitem{Huber:2019fxe}
T.~Huber, A.~von Manteuffel, E.~Panzer, R.~M. Schabinger, and G.~Yang, {\it
  {The Four-Loop Cusp Anomalous Dimension from the $\mathcal{N} = 4$ Sudakov
  Form Factor}},  \href{http://arxiv.org/abs/1912.13459}{{\tt
  arXiv:1912.13459}}.

\bibitem{Boels:2017skl}
R.~H. Boels, T.~Huber, and G.~Yang, {\it {Four-Loop Nonplanar Cusp Anomalous
  Dimension in N=4 Supersymmetric Yang-Mills Theory}},  {\em Phys. Rev. Lett.}
  {\bf 119} (2017), no.~20 201601, [\href{http://arxiv.org/abs/1705.03444}{{\tt
  arXiv:1705.03444}}].

\bibitem{Boels:2017ftb}
R.~H. Boels, T.~Huber, and G.~Yang, {\it {The Sudakov form factor at four loops
  in maximal super Yang-Mills theory}},  {\em JHEP} {\bf 01} (2018) 153,
  [\href{http://arxiv.org/abs/1711.08449}{{\tt arXiv:1711.08449}}].

\bibitem{Boels:2017fng}
R.~H. Boels, T.~Huber, and G.~Yang, {\it {The nonplanar cusp and collinear
  anomalous dimension at four loops in ${\mathcal N} = 4$ SYM theory}},  {\em
  PoS} {\bf RADCOR2017} (2017) 042,
  [\href{http://arxiv.org/abs/1712.07563}{{\tt arXiv:1712.07563}}].

\bibitem{Vogt:2004mw}
A.~Vogt, S.~Moch, and J.~A.~M. Vermaseren, {\it {The Three-loop splitting
  functions in QCD: The Singlet case}},  {\em Nucl. Phys.} {\bf B691} (2004)
  129--181, [\href{http://arxiv.org/abs/hep-ph/0404111}{{\tt hep-ph/0404111}}].

\bibitem{Moch:2004pa}
S.~Moch, J.~A.~M. Vermaseren, and A.~Vogt, {\it {The Three loop splitting
  functions in QCD: The Nonsinglet case}},  {\em Nucl. Phys.} {\bf B688} (2004)
  101--134, [\href{http://arxiv.org/abs/hep-ph/0403192}{{\tt hep-ph/0403192}}].

\bibitem{Gracey:1994nn}
J.~A. Gracey, {\it {Anomalous dimension of nonsinglet Wilson operators at O (1
  / N(f)) in deep inelastic scattering}},  {\em Phys. Lett.} {\bf B322} (1994)
  141--146, [\href{http://arxiv.org/abs/hep-ph/9401214}{{\tt hep-ph/9401214}}].

\bibitem{Davies:2016jie}
J.~Davies, A.~Vogt, B.~Ruijl, T.~Ueda, and J.~A.~M. Vermaseren, {\it
  {Large-$n_f$ contributions to the four-loop splitting functions in QCD}},
  {\em Nucl. Phys.} {\bf B915} (2017) 335--362,
  [\href{http://arxiv.org/abs/1610.07477}{{\tt arXiv:1610.07477}}].

\bibitem{Sudakov:1954sw}
V.~V. Sudakov, {\it {Vertex parts at very high-energies in quantum
  electrodynamics}},  {\em Sov. Phys. JETP} {\bf 3} (1956) 65--71. [Zh. Eksp.
  Teor. Fiz.30,87(1956)].

\bibitem{Collins:1980ih}
J.~C. Collins, {\it {Algorithm to Compute Corrections to the Sudakov
  Form-factor}},  {\em Phys. Rev.} {\bf D22} (1980) 1478.

\bibitem{Sen:1981sd}
A.~Sen, {\it {Asymptotic Behavior of the Sudakov Form-Factor in QCD}},  {\em
  Phys. Rev.} {\bf D24} (1981) 3281.

\bibitem{Korchemsky:1988hd}
G.~P. Korchemsky, {\it {Sudakov Form-factor in {QCD}}},  {\em Phys. Lett.} {\bf
  B220} (1989) 629--634.

\bibitem{Collins:1989bt}
J.~C. Collins, {\it {Sudakov form-factors}},  {\em Adv. Ser. Direct. High
  Energy Phys.} {\bf 5} (1989) 573--614,
  [\href{http://arxiv.org/abs/hep-ph/0312336}{{\tt hep-ph/0312336}}].

\bibitem{Magnea:1990zb}
L.~Magnea and G.~F. Sterman, {\it {Analytic continuation of the Sudakov
  form-factor in QCD}},  {\em Phys. Rev.} {\bf D42} (1990) 4222--4227.

\bibitem{Magnea:2000ss}
L.~Magnea, {\it {Analytic resummation for the quark form-factor in QCD}},  {\em
  Nucl. Phys.} {\bf B593} (2001) 269--288,
  [\href{http://arxiv.org/abs/hep-ph/0006255}{{\tt hep-ph/0006255}}].

\bibitem{Dixon:2008gr}
L.~J. Dixon, L.~Magnea, and G.~F. Sterman, {\it {Universal structure of
  subleading infrared poles in gauge theory amplitudes}},  {\em JHEP} {\bf 08}
  (2008) 022, [\href{http://arxiv.org/abs/0805.3515}{{\tt arXiv:0805.3515}}].

\bibitem{Falcioni:2019nxk}
G.~Falcioni, E.~Gardi, and C.~Milloy, {\it {Relating amplitude and PDF
  factorisation through Wilson-line geometries}},  {\em JHEP} {\bf 11} (2019)
  100, [\href{http://arxiv.org/abs/1909.00697}{{\tt arXiv:1909.00697}}].

\bibitem{Dixon:2017nat}
L.~J. Dixon, {\it {The Principle of Maximal Transcendentality and the Four-Loop
  Collinear Anomalous Dimension}},  \href{http://arxiv.org/abs/1712.07274}{{\tt
  arXiv:1712.07274}}.

\bibitem{Cachazo:2007ad}
F.~Cachazo, M.~Spradlin, and A.~Volovich, {\it {Four-Loop Collinear Anomalous
  Dimension in N = 4 Yang-Mills Theory}},  {\em Phys. Rev.} {\bf D76} (2007)
  106004, [\href{http://arxiv.org/abs/0707.1903}{{\tt arXiv:0707.1903}}].

\bibitem{vonManteuffel:2015gxa}
A.~von Manteuffel, E.~Panzer, and R.~M. Schabinger, {\it {On the Computation of
  Form Factors in Massless QCD with Finite Master Integrals}},  {\em Phys.
  Rev.} {\bf D93} (2016), no.~12 125014,
  [\href{http://arxiv.org/abs/1510.06758}{{\tt arXiv:1510.06758}}].

\bibitem{vonManteuffel:2019gpr}
A.~von Manteuffel and R.~M. Schabinger, {\it {Planar master integrals for
  four-loop form factors}},  {\em JHEP} {\bf 05} (2019) 073,
  [\href{http://arxiv.org/abs/1903.06171}{{\tt arXiv:1903.06171}}].

\bibitem{Ravindran:2004mb}
V.~Ravindran, J.~Smith, and W.~L. van Neerven, {\it {Two-loop corrections to
  Higgs boson production}},  {\em Nucl. Phys.} {\bf B704} (2005) 332--348,
  [\href{http://arxiv.org/abs/hep-ph/0408315}{{\tt hep-ph/0408315}}].

\bibitem{Manohar:2003vb}
A.~V. Manohar, {\it {Deep inelastic scattering as x ---\ensuremath{>} 1 using
  soft collinear effective theory}},  {\em Phys. Rev. D} {\bf 68} (2003)
  114019, [\href{http://arxiv.org/abs/hep-ph/0309176}{{\tt hep-ph/0309176}}].

\bibitem{Idilbi:2005ky}
A.~Idilbi and X.-d. Ji, {\it {Threshold resummation for Drell-Yan process in
  soft-collinear effective theory}},  {\em Phys. Rev. D} {\bf 72} (2005)
  054016, [\href{http://arxiv.org/abs/hep-ph/0501006}{{\tt hep-ph/0501006}}].

\bibitem{Idilbi:2005ni}
A.~Idilbi, X.-d. Ji, J.-P. Ma, and F.~Yuan, {\it {Threshold resummation for
  Higgs production in effective field theory}},  {\em Phys. Rev. D} {\bf 73}
  (2006) 077501, [\href{http://arxiv.org/abs/hep-ph/0509294}{{\tt
  hep-ph/0509294}}].

\bibitem{Becher:2007ty}
T.~Becher, M.~Neubert, and G.~Xu, {\it {Dynamical Threshold Enhancement and
  Resummation in Drell-Yan Production}},  {\em JHEP} {\bf 07} (2008) 030,
  [\href{http://arxiv.org/abs/0710.0680}{{\tt arXiv:0710.0680}}].

\bibitem{Moch:2005tm}
S.~Moch, J.~A.~M. Vermaseren, and A.~Vogt, {\it {Three-loop results for quark
  and gluon form-factors}},  {\em Phys. Lett.} {\bf B625} (2005) 245--252,
  [\href{http://arxiv.org/abs/hep-ph/0508055}{{\tt hep-ph/0508055}}].

\bibitem{Li:2014afw}
Y.~Li, A.~von Manteuffel, R.~M. Schabinger, and H.~X. Zhu, {\it {Soft-virtual
  corrections to Higgs production at N$^3$LO}},  {\em Phys. Rev.} {\bf D91}
  (2015) 036008, [\href{http://arxiv.org/abs/1412.2771}{{\tt
  arXiv:1412.2771}}].

\bibitem{Armoni:2006ux}
A.~Armoni, {\it {Anomalous Dimensions from a Spinning D5-Brane}},  {\em JHEP}
  {\bf 11} (2006) 009, [\href{http://arxiv.org/abs/hep-th/0608026}{{\tt
  hep-th/0608026}}].

\bibitem{Grozin:2017css}
A.~Grozin, J.~Henn, and M.~Stahlhofen, {\it {On the Casimir scaling violation
  in the cusp anomalous dimension at small angle}},  {\em JHEP} {\bf 10} (2017)
  052, [\href{http://arxiv.org/abs/1708.01221}{{\tt arXiv:1708.01221}}].

\bibitem{vanRitbergen:1997va}
T.~van Ritbergen, J.~A.~M. Vermaseren, and S.~A. Larin, {\it {The Four loop
  beta function in quantum chromodynamics}},  {\em Phys. Lett.} {\bf B400}
  (1997) 379--384, [\href{http://arxiv.org/abs/hep-ph/9701390}{{\tt
  hep-ph/9701390}}].

\bibitem{Becher:2019avh}
T.~Becher and M.~Neubert, {\it {Infrared Singularities of Scattering Amplitudes
  and N$^3$LL Resummation for $n$-Jet Processes}},
  \href{http://arxiv.org/abs/1908.11379}{{\tt arXiv:1908.11379}}.

\bibitem{Li:2016axz}
Y.~Li, D.~Neill, and H.~X. Zhu, {\it {An Exponential Regulator for Rapidity
  Divergences}},  {\em Submitted to: Phys. Rev. D} (2016)
  [\href{http://arxiv.org/abs/1604.00392}{{\tt arXiv:1604.00392}}].

\bibitem{Luo:2019szz}
M.-x. Luo, T.-Z. Yang, H.~X. Zhu, and Y.~J. Zhu, {\it {Quark Transverse Parton
  Distribution at the Next-to-Next-to-Next-to-Leading Order}},  {\em Phys. Rev.
  Lett.} {\bf 124} (2020), no.~9 092001,
  [\href{http://arxiv.org/abs/1912.05778}{{\tt arXiv:1912.05778}}].

\bibitem{Ebert:2020yqt}
M.~A. Ebert, B.~Mistlberger, and G.~Vita, {\it {Transverse momentum dependent
  PDFs at N$^3$LO}},  {\em JHEP} {\bf 09} (2020) 146,
  [\href{http://arxiv.org/abs/2006.05329}{{\tt arXiv:2006.05329}}].

\bibitem{Luo:2020epw}
M.-x. Luo, T.-Z. Yang, H.~X. Zhu, and Y.~J. Zhu, {\it {Unpolarized quark and
  gluon TMD PDFs and FFs at N$^{3}$LO}},  {\em JHEP} {\bf 06} (2021) 115,
  [\href{http://arxiv.org/abs/2012.03256}{{\tt arXiv:2012.03256}}].

\bibitem{Ebert:2020qef}
M.~A. Ebert, B.~Mistlberger, and G.~Vita, {\it {TMD fragmentation functions at
  N$^{3}$LO}},  {\em JHEP} {\bf 07} (2021) 121,
  [\href{http://arxiv.org/abs/2012.07853}{{\tt arXiv:2012.07853}}].

\bibitem{Ji:2004wu}
X.-d. Ji, J.-p. Ma, and F.~Yuan, {\it {QCD factorization for semi-inclusive
  deep-inelastic scattering at low transverse momentum}},  {\em Phys. Rev. D}
  {\bf 71} (2005) 034005, [\href{http://arxiv.org/abs/hep-ph/0404183}{{\tt
  hep-ph/0404183}}].

\bibitem{Collins:2011zzd}
J.~Collins, {\em {Foundations of perturbative QCD}}, vol.~32.
\newblock Cambridge University Press, 11, 2013.

\bibitem{Becher:2011dz}
T.~Becher and G.~Bell, {\it {Analytic Regularization in Soft-Collinear
  Effective Theory}},  {\em Phys. Lett. B} {\bf 713} (2012) 41--46,
  [\href{http://arxiv.org/abs/1112.3907}{{\tt arXiv:1112.3907}}].

\bibitem{Echevarria:2015byo}
M.~G. Echevarria, I.~Scimemi, and A.~Vladimirov, {\it {Universal transverse
  momentum dependent soft function at NNLO}},  {\em Phys. Rev. D} {\bf 93}
  (2016), no.~5 054004, [\href{http://arxiv.org/abs/1511.05590}{{\tt
  arXiv:1511.05590}}].

\bibitem{Ebert:2018gsn}
M.~A. Ebert, I.~Moult, I.~W. Stewart, F.~J. Tackmann, G.~Vita, and H.~X. Zhu,
  {\it {Subleading power rapidity divergences and power corrections for
  q$_{T}$}},  {\em JHEP} {\bf 04} (2019) 123,
  [\href{http://arxiv.org/abs/1812.08189}{{\tt arXiv:1812.08189}}].

\bibitem{Vladimirov:2020umg}
A.~A. Vladimirov, {\it {Self-contained definition of Collins-Soper kernel}},
  \href{http://arxiv.org/abs/2003.02288}{{\tt arXiv:2003.02288}}.

\bibitem{Zhu:2020ftr}
Y.~J. Zhu, {\it {Double soft current at one-loop in QCD}},
  \href{http://arxiv.org/abs/2009.08919}{{\tt arXiv:2009.08919}}.

\bibitem{Catani:2021kcy}
S.~Catani and L.~Cieri, {\it {Multiple soft radiation at one-loop order and the
  emission of a soft quark\textendash{}antiquark pair}},  {\em Eur. Phys. J. C}
  {\bf 82} (2022), no.~2 97, [\href{http://arxiv.org/abs/2108.13309}{{\tt
  arXiv:2108.13309}}].

\bibitem{Braun:2018mxm}
V.~M. Braun, A.~N. Manashov, S.~O. Moch, and M.~Strohmaier, {\it {Conformal
  symmetry of QCD in $d$-dimensions}},  {\em Phys. Lett.} {\bf B793} (2019)
  78--84, [\href{http://arxiv.org/abs/1810.04993}{{\tt arXiv:1810.04993}}].

\bibitem{Collins:1988ig}
J.~C. Collins, D.~E. Soper, and G.~F. Sterman, {\it {Soft Gluons and
  Factorization}},  {\em Nucl. Phys.} {\bf B308} (1988) 833--856.

\bibitem{Ebert:2020sfi}
M.~A. Ebert, B.~Mistlberger, and G.~Vita, {\it {The Energy-Energy Correlation
  in the back-to-back limit at N$^{3}$LO and N$^{3}$LL'}},  {\em JHEP} {\bf 08}
  (2021) 022, [\href{http://arxiv.org/abs/2012.07859}{{\tt arXiv:2012.07859}}].

\bibitem{Gao:2019ojf}
A.~Gao, H.~T. Li, I.~Moult, and H.~X. Zhu, {\it {The Transverse Energy-Energy
  Correlator in the Back-to-Back Limit}},
  \href{http://arxiv.org/abs/1901.04497}{{\tt arXiv:1901.04497}}.

\bibitem{Moult:2019vou}
I.~Moult, G.~Vita, and K.~Yan, {\it {Subleading power resummation of rapidity
  logarithms: the energy-energy correlator in $ \mathcal{N} $ = 4 SYM}},  {\em
  JHEP} {\bf 07} (2020) 005, [\href{http://arxiv.org/abs/1912.02188}{{\tt
  arXiv:1912.02188}}].

\bibitem{Li:2021txc}
H.~T. Li, Y.~Makris, and I.~Vitev, {\it {Energy-energy correlators in Deep
  Inelastic Scattering}},  \href{http://arxiv.org/abs/2102.05669}{{\tt
  arXiv:2102.05669}}.

\bibitem{Li:2020bub}
H.~T. Li, I.~Vitev, and Y.~J. Zhu, {\it {Transverse-Energy-Energy Correlations
  in Deep Inelastic Scattering}},  {\em JHEP} {\bf 11} (2020) 051,
  [\href{http://arxiv.org/abs/2006.02437}{{\tt arXiv:2006.02437}}].

\bibitem{Hofman:2008ar}
D.~M. Hofman and J.~Maldacena, {\it {Conformal collider physics: Energy and
  charge correlations}},  {\em JHEP} {\bf 05} (2008) 012,
  [\href{http://arxiv.org/abs/0803.1467}{{\tt arXiv:0803.1467}}].

\bibitem{Belitsky:2013bja}
A.~V. Belitsky, S.~Hohenegger, G.~P. Korchemsky, E.~Sokatchev, and
  A.~Zhiboedov, {\it {Event shapes in $\mathcal{N} = 4$ super-Yang-Mills
  theory}},  {\em Nucl. Phys.} {\bf B884} (2014) 206--256,
  [\href{http://arxiv.org/abs/1309.1424}{{\tt arXiv:1309.1424}}].

\bibitem{Belitsky:2013xxa}
A.~V. Belitsky, S.~Hohenegger, G.~P. Korchemsky, E.~Sokatchev, and
  A.~Zhiboedov, {\it {From correlation functions to event shapes}},  {\em Nucl.
  Phys.} {\bf B884} (2014) 305--343,
  [\href{http://arxiv.org/abs/1309.0769}{{\tt arXiv:1309.0769}}].

\bibitem{Sveshnikov:1995vi}
N.~A. Sveshnikov and F.~V. Tkachov, {\it {Jets and quantum field theory}},
  {\em Phys. Lett.} {\bf B382} (1996) 403--408,
  [\href{http://arxiv.org/abs/hep-ph/9512370}{{\tt hep-ph/9512370}}].

\bibitem{Korchemsky:1999kt}
G.~P. Korchemsky and G.~F. Sterman, {\it {Power corrections to event shapes and
  factorization}},  {\em Nucl. Phys.} {\bf B555} (1999) 335--351,
  [\href{http://arxiv.org/abs/hep-ph/9902341}{{\tt hep-ph/9902341}}].

\bibitem{Lee:2006nr}
C.~Lee and G.~F. Sterman, {\it {Momentum Flow Correlations from Event Shapes:
  Factorized Soft Gluons and Soft-Collinear Effective Theory}},  {\em Phys.
  Rev.} {\bf D75} (2007) 014022,
  [\href{http://arxiv.org/abs/hep-ph/0611061}{{\tt hep-ph/0611061}}].

\bibitem{Belitsky:2013ofa}
A.~V. Belitsky, S.~Hohenegger, G.~P. Korchemsky, E.~Sokatchev, and
  A.~Zhiboedov, {\it {Energy-Energy Correlations in N=4 Supersymmetric
  Yang-Mills Theory}},  {\em Phys. Rev. Lett.} {\bf 112} (2014), no.~7 071601,
  [\href{http://arxiv.org/abs/1311.6800}{{\tt arXiv:1311.6800}}].

\bibitem{Alday:2010zy}
L.~F. Alday, B.~Eden, G.~P. Korchemsky, J.~Maldacena, and E.~Sokatchev, {\it
  {From correlation functions to Wilson loops}},  {\em JHEP} {\bf 09} (2011)
  123, [\href{http://arxiv.org/abs/1007.3243}{{\tt arXiv:1007.3243}}].

\bibitem{Alday:2013cwa}
L.~F. Alday and A.~Bissi, {\it {Higher-spin correlators}},  {\em JHEP} {\bf 10}
  (2013) 202, [\href{http://arxiv.org/abs/1305.4604}{{\tt arXiv:1305.4604}}].

\bibitem{Basso:2006nk}
B.~Basso and G.~Korchemsky, {\it {Anomalous dimensions of high-spin operators
  beyond the leading order}},  {\em Nucl. Phys. B} {\bf 775} (2007) 1--30,
  [\href{http://arxiv.org/abs/hep-th/0612247}{{\tt hep-th/0612247}}].

\bibitem{Chen:2020uvt}
H.~Chen, T.-Z. Yang, H.~X. Zhu, and Y.~J. Zhu, {\it {Analytic Continuation and
  Reciprocity Relation for Collinear Splitting in QCD}},  {\em Chin. Phys. C}
  {\bf 45} (2021), no.~4 043101, [\href{http://arxiv.org/abs/2006.10534}{{\tt
  arXiv:2006.10534}}].

\bibitem{Kologlu:2019mfz}
M.~Kologlu, P.~Kravchuk, D.~Simmons-Duffin, and A.~Zhiboedov, {\it {The
  light-ray OPE and conformal colliders}},  {\em JHEP} {\bf 01} (2021) 128,
  [\href{http://arxiv.org/abs/1905.01311}{{\tt arXiv:1905.01311}}].

\bibitem{Dixon:2019uzg}
L.~J. Dixon, I.~Moult, and H.~X. Zhu, {\it {Collinear limit of the
  energy-energy correlator}},  {\em Phys. Rev.} {\bf D100} (2019), no.~1
  014009, [\href{http://arxiv.org/abs/1905.01310}{{\tt arXiv:1905.01310}}].

\bibitem{Chen:2020vvp}
H.~Chen, I.~Moult, X.~Zhang, and H.~X. Zhu, {\it {Rethinking jets with energy
  correlators: Tracks, resummation, and analytic continuation}},  {\em Phys.
  Rev. D} {\bf 102} (2020), no.~5 054012,
  [\href{http://arxiv.org/abs/2004.11381}{{\tt arXiv:2004.11381}}].

\bibitem{Zhu:2014fma}
H.~X. Zhu, {\it {On the calculation of soft phase space integral}},  {\em JHEP}
  {\bf 02} (2015) 155, [\href{http://arxiv.org/abs/1501.00236}{{\tt
  arXiv:1501.00236}}].

\bibitem{Banfi:2018mcq}
A.~Banfi, B.~K. El-Menoufi, and P.~F. Monni, {\it {The Sudakov radiator for jet
  observables and the soft physical coupling}},  {\em JHEP} {\bf 01} (2019)
  083, [\href{http://arxiv.org/abs/1807.11487}{{\tt arXiv:1807.11487}}].

\bibitem{Gatheral:1983cz}
J.~G.~M. Gatheral, {\it {Exponentiation of Eikonal Cross-sections in Nonabelian
  Gauge Theories}},  {\em Phys. Lett.} {\bf 133B} (1983) 90--94.

\bibitem{Frenkel:1984pz}
J.~Frenkel and J.~C. Taylor, {\it {NONABELIAN EIKONAL EXPONENTIATION}},  {\em
  Nucl. Phys.} {\bf B246} (1984) 231--245.

\bibitem{Li:2014bfa}
Y.~Li, A.~von Manteuffel, R.~M. Schabinger, and H.~X. Zhu, {\it {N$^3$LO Higgs
  boson and Drell-Yan production at threshold: The one-loop two-emission
  contribution}},  {\em Phys. Rev.} {\bf D90} (2014), no.~5 053006,
  [\href{http://arxiv.org/abs/1404.5839}{{\tt arXiv:1404.5839}}].

\bibitem{Li:2013lsa}
Y.~Li and H.~X. Zhu, {\it {Single soft gluon emission at two loops}},  {\em
  JHEP} {\bf 11} (2013) 080, [\href{http://arxiv.org/abs/1309.4391}{{\tt
  arXiv:1309.4391}}].

\bibitem{Duhr:2013msa}
C.~Duhr and T.~Gehrmann, {\it {The two-loop soft current in dimensional
  regularization}},  {\em Phys. Lett. B} {\bf 727} (2013) 452--455,
  [\href{http://arxiv.org/abs/1309.4393}{{\tt arXiv:1309.4393}}].

\bibitem{Nogueira:1991ex}
P.~Nogueira, {\it {Automatic Feynman graph generation}},  {\em J. Comput.
  Phys.} {\bf 105} (1993) 279--289.

\bibitem{Catani:2019nqv}
S.~Catani, D.~Colferai, and A.~Torrini, {\it {Triple (and quadruple) soft-gluon
  radiation in QCD hard scattering}},  {\em JHEP} {\bf 01} (2020) 118,
  [\href{http://arxiv.org/abs/1908.01616}{{\tt arXiv:1908.01616}}].

\bibitem{Anastasiou:2013srw}
C.~Anastasiou, C.~Duhr, F.~Dulat, and B.~Mistlberger, {\it {Soft triple-real
  radiation for Higgs production at N3LO}},  {\em JHEP} {\bf 07} (2013) 003,
  [\href{http://arxiv.org/abs/1302.4379}{{\tt arXiv:1302.4379}}].

\bibitem{Anastasiou:2014vaa}
C.~Anastasiou, C.~Duhr, F.~Dulat, E.~Furlan, T.~Gehrmann, F.~Herzog, and
  B.~Mistlberger, {\it {Higgs boson gluon\textendash{}fusion production at
  threshold in N$^3$LO QCD}},  {\em Phys. Lett. B} {\bf 737} (2014) 325--328,
  [\href{http://arxiv.org/abs/1403.4616}{{\tt arXiv:1403.4616}}].

\bibitem{Anastasiou:2015yha}
C.~Anastasiou, C.~Duhr, F.~Dulat, E.~Furlan, F.~Herzog, and B.~Mistlberger,
  {\it {Soft expansion of double-real-virtual corrections to Higgs production
  at N$^{3}$LO}},  {\em JHEP} {\bf 08} (2015) 051,
  [\href{http://arxiv.org/abs/1505.04110}{{\tt arXiv:1505.04110}}].

\bibitem{Ruijl:2017dtg}
B.~Ruijl, T.~Ueda, and J.~Vermaseren, {\it {FORM version 4.2}},
  \href{http://arxiv.org/abs/1707.06453}{{\tt arXiv:1707.06453}}.

\bibitem{Anastasiou:2002yz}
C.~Anastasiou and K.~Melnikov, {\it {Higgs boson production at hadron colliders
  in NNLO QCD}},  {\em Nucl. Phys. B} {\bf 646} (2002) 220--256,
  [\href{http://arxiv.org/abs/hep-ph/0207004}{{\tt hep-ph/0207004}}].

\bibitem{Chetyrkin:1981qh}
K.~G. Chetyrkin and F.~V. Tkachov, {\it {Integration by Parts: The Algorithm to
  Calculate beta Functions in 4 Loops}},  {\em Nucl. Phys.} {\bf B192} (1981)
  159--204.

\bibitem{Lee:2012cn}
R.~N. Lee, {\it {Presenting LiteRed: a tool for the Loop InTEgrals REDuction}},
   \href{http://arxiv.org/abs/1212.2685}{{\tt arXiv:1212.2685}}.

\bibitem{Kotikov:1990kg}
A.~V. Kotikov, {\it {Differential equations method: New technique for massive
  Feynman diagrams calculation}},  {\em Phys. Lett. B} {\bf 254} (1991)
  158--164.

\bibitem{Henn:2013pwa}
J.~M. Henn, {\it {Multiloop integrals in dimensional regularization made
  simple}},  {\em Phys. Rev. Lett.} {\bf 110} (2013) 251601,
  [\href{http://arxiv.org/abs/1304.1806}{{\tt arXiv:1304.1806}}].

\bibitem{Meyer:2016slj}
C.~Meyer, {\it {Transforming differential equations of multi-loop Feynman
  integrals into canonical form}},  {\em JHEP} {\bf 04} (2017) 006,
  [\href{http://arxiv.org/abs/1611.01087}{{\tt arXiv:1611.01087}}].

\bibitem{Ji:2019ewn}
X.~Ji, Y.~Liu, and Y.-S. Liu, {\it {Transverse-momentum-dependent parton
  distribution functions from large-momentum effective theory}},  {\em Phys.
  Lett. B} {\bf 811} (2020) 135946,
  [\href{http://arxiv.org/abs/1911.03840}{{\tt arXiv:1911.03840}}].

\bibitem{Ebert:2018gzl}
M.~A. Ebert, I.~W. Stewart, and Y.~Zhao, {\it {Determining the Nonperturbative
  Collins-Soper Kernel From Lattice QCD}},  {\em Phys. Rev. D} {\bf 99} (2019),
  no.~3 034505, [\href{http://arxiv.org/abs/1811.00026}{{\tt
  arXiv:1811.00026}}].

\bibitem{Ebert:2019tvc}
M.~A. Ebert, I.~W. Stewart, and Y.~Zhao, {\it {Renormalization and Matching for
  the Collins-Soper Kernel from Lattice QCD}},  {\em JHEP} {\bf 03} (2020) 099,
  [\href{http://arxiv.org/abs/1910.08569}{{\tt arXiv:1910.08569}}].

\bibitem{Shanahan:2020zxr}
P.~Shanahan, M.~Wagman, and Y.~Zhao, {\it {Collins-Soper kernel for TMD
  evolution from lattice QCD}},  {\em Phys. Rev. D} {\bf 102} (2020), no.~1
  014511, [\href{http://arxiv.org/abs/2003.06063}{{\tt arXiv:2003.06063}}].

\bibitem{LatticeParton:2020uhz}
{\bf Lattice Parton} Collaboration, Q.-A. Zhang et~al., {\it {Lattice QCD
  Calculations of Transverse-Momentum-Dependent Soft Function through
  Large-Momentum Effective Theory}},  {\em Phys. Rev. Lett.} {\bf 125} (2020),
  no.~19 192001, [\href{http://arxiv.org/abs/2005.14572}{{\tt
  arXiv:2005.14572}}].

\bibitem{Li:2021wvl}
Y.~Li et~al., {\it {Lattice QCD Study of Transverse-Momentum Dependent Soft
  Function}},  {\em Phys. Rev. Lett.} {\bf 128} (2022), no.~6 062002,
  [\href{http://arxiv.org/abs/2106.13027}{{\tt arXiv:2106.13027}}].

\bibitem{Almelid:2015jia}
{\O}.~Almelid, C.~Duhr, and E.~Gardi, {\it {Three-loop corrections to the soft
  anomalous dimension in multileg scattering}},  {\em Phys. Rev. Lett.} {\bf
  117} (2016), no.~17 172002, [\href{http://arxiv.org/abs/1507.00047}{{\tt
  arXiv:1507.00047}}].

\bibitem{Almelid:2017qju}
{\O}.~Almelid, C.~Duhr, E.~Gardi, A.~McLeod, and C.~D. White, {\it
  {Bootstrapping the QCD soft anomalous dimension}},  {\em JHEP} {\bf 09}
  (2017) 073, [\href{http://arxiv.org/abs/1706.10162}{{\tt arXiv:1706.10162}}].

\bibitem{Fadin:1995xg}
V.~S. Fadin, M.~I. Kotsky, and R.~Fiore, {\it {Gluon Reggeization in QCD in the
  next-to-leading order}},  {\em Phys. Lett.} {\bf B359} (1995) 181--188.

\bibitem{Fadin:1995km}
V.~S. Fadin, R.~Fiore, and A.~Quartarolo, {\it {Reggeization of quark quark
  scattering amplitude in QCD}},  {\em Phys. Rev.} {\bf D53} (1996) 2729--2741,
  [\href{http://arxiv.org/abs/hep-ph/9506432}{{\tt hep-ph/9506432}}].

\bibitem{Fadin:1996tb}
V.~S. Fadin, R.~Fiore, and M.~I. Kotsky, {\it {Gluon Regge trajectory in the
  two loop approximation}},  {\em Phys. Lett.} {\bf B387} (1996) 593--602,
  [\href{http://arxiv.org/abs/hep-ph/9605357}{{\tt hep-ph/9605357}}].

\bibitem{Korchemskaya:1996je}
I.~A. Korchemskaya and G.~P. Korchemsky, {\it {Evolution equation for gluon
  Regge trajectory}},  {\em Phys. Lett.} {\bf B387} (1996) 346--354,
  [\href{http://arxiv.org/abs/hep-ph/9607229}{{\tt hep-ph/9607229}}].

\bibitem{Blumlein:1998ib}
J.~Blumlein, V.~Ravindran, and W.~L. van Neerven, {\it {On the gluon Regge
  trajectory in O alpha-s**2}},  {\em Phys. Rev.} {\bf D58} (1998) 091502,
  [\href{http://arxiv.org/abs/hep-ph/9806357}{{\tt hep-ph/9806357}}].

\bibitem{DelDuca:2001gu}
V.~Del~Duca and E.~W.~N. Glover, {\it {The High-energy limit of QCD at two
  loops}},  {\em JHEP} {\bf 10} (2001) 035,
  [\href{http://arxiv.org/abs/hep-ph/0109028}{{\tt hep-ph/0109028}}].

\bibitem{Bogdan:2002sr}
A.~V. Bogdan, V.~Del~Duca, V.~S. Fadin, and E.~W.~N. Glover, {\it {The Quark
  Regge trajectory at two loops}},  {\em JHEP} {\bf 03} (2002) 032,
  [\href{http://arxiv.org/abs/hep-ph/0201240}{{\tt hep-ph/0201240}}].

\bibitem{Caola:2021izf}
F.~Caola, A.~Chakraborty, G.~Gambuti, A.~von Manteuffel, and L.~Tancredi, {\it
  {Three-loop gluon scattering in QCD and the gluon Regge trajectory}},
  \href{http://arxiv.org/abs/2112.11097}{{\tt arXiv:2112.11097}}.

\bibitem{DelDuca:2021vjq}
V.~Del~Duca, R.~Marzucca, and B.~Verbeek, {\it {The gluon Regge trajectory at
  three loops from planar Yang-Mills theory}},  {\em JHEP} {\bf 01} (2022) 149,
  [\href{http://arxiv.org/abs/2111.14265}{{\tt arXiv:2111.14265}}].

\bibitem{Falcioni:2021dgr}
G.~Falcioni, E.~Gardi, N.~Maher, C.~Milloy, and L.~Vernazza, {\it
  {Disentangling the Regge Cut and Regge Pole in Perturbative QCD}},  {\em
  Phys. Rev. Lett.} {\bf 128} (2022), no.~13 132001,
  [\href{http://arxiv.org/abs/2112.11098}{{\tt arXiv:2112.11098}}].

\bibitem{Falcioni:2021buo}
G.~Falcioni, E.~Gardi, N.~Maher, C.~Milloy, and L.~Vernazza, {\it {Scattering
  amplitudes in the Regge limit and the soft anomalous dimension through four
  loops}},  {\em JHEP} {\bf 03} (2022) 053,
  [\href{http://arxiv.org/abs/2111.10664}{{\tt arXiv:2111.10664}}].

\bibitem{Duhr:2022yyp}
C.~Duhr, B.~Mistlberger, and G.~Vita, {\it {The Four-Loop Rapidity Anomalous
  Dimension and Event Shapes to Fourth Logarithmic Order}},
  \href{http://arxiv.org/abs/2205.02242}{{\tt arXiv:2205.02242}}.

\end{thebibliography}\endgroup

\end{document}